\documentclass[twocolumn,tighten]{aa} 

\usepackage{natbib}
\usepackage{graphicx}		
\usepackage{url}
\usepackage{epstopdf}
\usepackage{color, xcolor}
\usepackage{multirow}
\usepackage{ragged2e}
\usepackage{hyperref}

\hypersetup{
colorlinks, 
allcolors = {blue!50!black}
}

\usepackage{amsmath} 
\usepackage{mathtools}
\usepackage{booktabs}
\usepackage{comment}
\usepackage{float}
\usepackage{sidecap}
\usepackage{wrapfig}
\usepackage{ulem}
\usepackage{footmisc}
\usepackage{array}
\usepackage{capt-of}
\usepackage{orcidlink}
\usepackage[varg]{txfonts}

\DeclareGraphicsExtensions{.pdf,.png,.jpeg}
\def\lsim{\mathrel{\raise.3ex\hbox{$<$\kern-.75em\lower1ex\hbox{$\sim$}}}}
\def\gsim{\mathrel{\raise.3ex\hbox{$>$\kern-.75em\lower1ex\hbox{$\sim$}}}}
\def\gtwid{\mathrel{\raise.3ex\hbox{$>$\kern-.75em\lower1ex\hbox{$\sim$}}}}
\def\proptwid{\mathrel{\raise.3ex\hbox{$\propto$\kern-.75em\lower1ex\hbox{$\sim$}}}}

\definecolor{orange}{rgb}{1,0.5,0}

\newcommand\oj{OJ\,287\xspace}
\newcommand\ra{RadioAstron\xspace}

\newcommand\difmap{{\tt Difmap}\xspace}
\newcommand\dearth{$D_{\oplus}$}
\newcommand\tb{$T_{\rm B}$\xspace}
\newcommand\srt{{SRT}\xspace}

\newcommand{\rev}[1]{#1}  

\begin{document}

\title{Unveiling the Bent Jet Structure and Polarization of \oj at 1.7\,GHz with Space VLBI}

\author{
Ilje Cho\inst{\ref{iaa},\ref{kasi},\ref{yonsei}}\orcidlink{0000-0001-6083-7521} \and  
Jos\'e L. G\'omez\inst{\ref{iaa}}\orcidlink{0000-0003-4190-7613} \and
Rocco Lico\inst{\ref{iaa},\ref{inaf}}\orcidlink{0000-0001-7361-2460} \and
Guang-Yao Zhao\inst{\ref{iaa},\ref{mpifr}}\orcidlink{0000-0002-4417-1659} \and
Efthalia Traianou\inst{\ref{iaa}}\orcidlink{0000-0002-1209-650} \and
Rohan Dahale\inst{\ref{iaa}}\orcidlink{0000-0001-6982-9034} \and
Antonio Fuentes\inst{\ref{iaa}}\orcidlink{0000-0002-8773-4933} \and
Teresa Toscano\inst{\ref{iaa}}\orcidlink{0000-0003-3658-7862} \and
Marianna Foschi\inst{\ref{iaa}}\orcidlink{0000-0001-8147-4993} \and
Yuri Y. Kovalev\inst{\ref{mpifr},\ref{lebedev},\ref{mipt}}\orcidlink{0000-0001-9303-3263} \and
Andrei Lobanov\inst{\ref{mpifr},\ref{mipt}}\orcidlink{0000-0003-1622-1484} \and
Alexander B. Pushkarev\inst{\ref{crao},\ref{lebedev}}\orcidlink{0000-0002-9702-2307} \and
Leonid~I.~Gurvits\inst{\ref{jive},\ref{delft}}\orcidlink{0000-0002-0694-2459} \and
Jae-Young Kim\inst{\ref{knu},\ref{mpifr}}\orcidlink{0000-0001-8229-7183} \and
Mikhail Lisakov\inst{\ref{lebedev}}\orcidlink{0000-0001-6088-3819} \and
Petr Voitsik\inst{\ref{lebedev}}\orcidlink{0000-0002-1290-1629} \and
Ioannis Myserlis\inst{\ref{iram},\ref{mpifr}}\orcidlink{0000-0003-3025-9497} \and
Felix P\"otzl\inst{\ref{forth},\ref{mpifr}}\orcidlink{0000-0002-6579-8311} \and
Eduardo Ros\inst{\ref{mpifr}}\orcidlink{0000-0001-9503-4892} 
} 

\institute{
Instituto de Astrof\'{\i}sica de Andaluc\'{\i}a-CSIC, Glorieta de la Astronom\'{\i}a s/n, E-18008 Granada, Spain \label{iaa} \\ 
\email{icho@kasi.re.kr} 
\and 
%
Korea Astronomy and Space Science Institute, Daedeok-daero 776, Yuseong-gu, Daejeon 34055, Republic of Korea \label{kasi}
\and 
Department of Astronomy, Yonsei University, Yonsei-ro 50, Seodaemun-gu, 03722 Seoul, Republic of Korea \label{yonsei}
\and 
%
INAF -- Istituto di Radioastronomia, Via P. Gobetti 101, I-40129 Bologna, Italy \label{inaf} \and
Max-Planck-Institut f\"ur Radioastronomie, Auf dem H\"ugel 69, D-53121 Bonn, Germany \label{mpifr} \and 
%
Lebedev Physical Institute of the Russian Academy of Sciences, Leninsky prospekt 53, 119991 Moscow, Russia \label{lebedev} \and
Moscow Institute of Physics and Technology, Institutsky per. 9, Dolgoprudny, Moscow region, 141700, Russia \label{mipt} \and 
Crimean Astrophysical Observatory, 98409 Nauchny, Crimea, Russia \label{crao} \and
%
Joint Institute for VLBI ERIC (JIVE), Oude Hoogeveensedijk 4, 7991 PD Dwingeloo, The Netherlands \label{jive} \and
Faculty of Aerospace Engineering, Delft University of Technology, Kluyverweg 1, 2629 HS Delft, The Netherlands \label{delft} \and
%
Department of Astronomy and Atmospheric Sciences, Kyungpook National University, Daegu 702-701, Republic of Korea \label{knu} \and 
%
Institut de Radioastronomie Millim\'{e}trique, Avenida Divina Pastora, 7, Local 20, E18012 Granada, Spain \label{iram} \and
%
Institute of Astrophysics, Foundation for Research and Technology – Hellas, N. Plastira 100, Voutes GR-70013, Heraklion, Greece \label{forth} 
} 

\date{December 8, 2023}

\abstract {
We present total intensity and linear polarization images of \oj at 1.68\,GHz, obtained through space-based very long baseline interferometry (VLBI) observations with \ra on April 16, 2016. The observations were conducted using a ground array consisting of the Very Long Baseline Array (VLBA) and the European VLBI Network (EVN). Ground-space fringes were detected with a maximum projected baseline length of $\sim\!5.6\,$Earth's diameter, resulting in an angular resolution of $\sim\!530\,\mu$as. With this unprecedented resolution at such a low frequency, the progressively bending jet structure of \oj has been resolved up to $\sim10\,$parsec of the projected distance from the radio core. In comparison with close-in-time VLBI observations at 15, 43, 86\,GHz from MOJAVE and VLBA-BU-BLAZAR monitoring projects, we obtain the spectral index map showing the opaque core and optically thin jet components. The optically thick core has a brightness temperature of $\sim10^{13}\,$K, and is further resolved into two sub-components at higher frequencies labeled C1 and C2. These sub-components exhibit a transition from optically thick to thin, with a synchrotron self-absorption (SSA) turnover frequency estimated to be $\sim\!33$ and $\sim\!11.5$\,GHz, and a turnover flux density $\sim\!4$ and $\sim\!0.7$\,Jy, respectively. 
Assuming a Doppler boosting factor of 10, the SSA values provide the estimate of the magnetic field strengths from SSA of $\sim\!3.4\,$G for C1 and $\sim\!1.0\,$G for C2. The magnetic field strengths assuming equipartition arguments are also estimated as $\sim\!2.6\,$G and $\sim\!1.6\,$G, respectively. The integrated degree of linear polarization is found to be approximately $\sim$2.5\,\%, with the electric vector position angle being well aligned with the local jet direction at the core region. This alignment suggests a predominant toroidal magnetic field, which is in agreement with the jet formation model that requires a helical magnetic field anchored to either the black hole ergosphere or the accretion disk. Further downstream, the jet seems to be predominantly threaded by a poloidal magnetic field. 
} 


\maketitle
%


\section{Introduction} \label{sec:intro}

\oj (0851$+$202, J0854$+$2005, $z=0.306$; \citealt{Stickel_1989}) is one of the most studied BL Lacertae objects and considered to be a strong candidate for harboring a supermassive black hole binary (SMBHB) system at its center \citep[e.g.,][]{Sillanpaa_1988, Lehto_1996}. For that reason, it attracts special attention as one of the most promising gravitational wave emitters at nano-hertz frequencies \citep[e.g.,][]{Valtonen_2021}. In the optical regime, two quasiperiodic outbursts of $\sim12\,$yrs and $\sim60\,$yrs have been observed, which may be attributed to the gravitationally bound system of SMBHB \citep[e.g.,][]{Valtonen_2016,Komossa_2020}. Accounting for general relativistic effects, it is predicted that the system consists of primary and secondary supermassive black holes (SMBHs) of $\sim1.8\times10^{10}\,{\rm M}_\odot$ and $\sim1.5\times10^{8}\,{\rm M}_\odot$, respectively \citep[e.g.,][]{Dey_2018}, with an orbital major axis of 0.112\,pc ($\sim26\,$microarcseconds, $\mu$as; e.g., \citealt{Valtonen_2008}). However, recently, \citet{Komossa_2023} suggested an alternative SMBHB model which involves a primary SMBH of a much lower mass, $10^{8}\,{\rm M}_\odot$, and does not require strong orbital precession. 
In radio regime, \oj has been intensively studied, especially by means of very long baseline interferometry (VLBI) observations. Owing to its massive central black hole and low redshift, this source has been a very efficient study case for investigating the launching and acceleration mechanism of relativistic jets in active galactic nuclei (AGN) by resolving its innermost structure. Until now, the formation of a jet has been explained by two different mechanisms: 1) by the rotational energy of the black hole (BZ process; \citealt{Blandford_1977}), and 2) by the differential rotation of the accretion flow (BP process; \citealt{Blandford_1982}). In addition the combination of both has been also suggested \citep[e.g.,][]{Chiaberge_2000}. To probe these mechanisms, it is essential to resolve the fine structure in the vicinity of the black hole, where the jet base is thought to be located. This necessitates observations with extremely high angular resolution. An example of such study has been presented recently by the Event Horizon Telescope (EHT). In combination with multi-wavelength observations by other instruments, the EHT has revealed a potential prevalence of the BZ process in the center of M\,87 by resolving its black hole shadow at 230\,GHz with angular resolution of $\sim 20$\,$\mu$as \citep[e.g.,][]{ehtc2019a}. A similar EHT study of \oj will be presented in a forthcoming paper (J.~L.~G{\'o}mez et al. in prep.). 

Interestingly, many previous studies have found the jet of \oj to be wobbling on parsec (pc) scales, as inferred from the observation of position angle (PA) variations with a period $\sim$24--30\,yrs \citep[e.g.,][]{Cohen_2017, Britzen_2018}. This might support the SMBHB scenario, involving the jet precession by the binary companion \citep[e.g.,][]{Dey_2021}. However, alternative models cannot still be ruled out, such as the accretion flow instabilities \citep{Agudo_2012} or resonant accretion of magnetic field lines \citep{Villforth_2010}, Doppler beaming effects of a precessing jet \citep{Britzen_2018} or jet helicity \citep{Butuzova_2020}, Lense-Thirring effect by the misalignment of black hole and accretion disk spin axis \citep[e.g.,][]{Lense_1918, Thirring_1918, Liska_2018, Chatterjee_2020}, and a helical distortions of the jet by magneto-hydro-dynamic (MHD) instabilities \citep[e.g., current-driven or Kelvin–Helmholtz;][]{Mizuno_2012, Perucho_2012, Vega-Garcia_2019}. 

It has been also shown that the jet in \oj is bent at de-projected distances from the central black hole of several to tens parsecs. While the innermost jet is mostly extended along the northwest direction \citep[e.g.,][]{Zhao_2022}, the progressive jet bending towards southwest is found downstream the jet \citep[e.g.,][]{Gomez_2022}. 
This might be an indication of the change of the jet launching PA with time ``imprinted'' in the PA of different sections of the downstream jet. Recently, \citet{Lico_2022} have indeed found that the jet structure changes at 86\,GHz over the period of $\sim2\,$yrs, 2015$-$2017. These changes can also be investigated in the downstream jet structure through observations at lower frequencies (e.g., $\lesssim\,1\,$GHz). 

The \ra space VLBI mission \citep{Kardashev_2013} offered an excellent opportunity to probe larger scale jet structures with sub-milliarcsecond (mas) angular resolution. 
The main component of the mission was a 10-m radio telescope on board the Spektr-R satellite which operated on a highly eccentric orbit with the initial period of $\sim9\,$days. This orbit near the apogee provided the longest interferometric baselines to ground radio telescopes of about 350,000\,km. The spaceborne radio telescope (\srt) of this mission was able to observe at four frequency bands (0.32, 1.6, 4.8, and 22\,GHz) in the dual polarization mode. A brief comparison of the specifications of the \ra mission with two other space VLBI systems, TDRSS\footnote{the geostationary Tracking and Data Relay Satellite System} and VSOP\footnote{the VLBI Space Observatory Program}, is given in \citet{Gurvits_2023}. With global VLBI arrays (e.g., the Very Long Baseline Array, VLBA, and the European VLBI Network, EVN) as ground-based elements of the system, the \ra observations could provide interferometric detections on baselines reaching several times the Earth diameter, proportionally exceeding the highest angular resolution accessible to ground-only VLBI networks. 

In this study, we have analyzed the \ra observation of \oj at 1.68\,GHz with dual polarization mode. The layout of this paper is as follows. We first summarize the \ra observation, data reduction and imaging processes in \autoref{sec:obs_calibration}. In \autoref{sec:results}, the results from total intensity and polarization images are presented. Here the independent results from adjacent VLBA observations at 15 and 43\,GHz are compared together to derive the spectral index and light curves. In \autoref{sec:summary}, we further discuss the results and summarize their implications for future studies. The following cosmological parameters are adopted throughout this paper: $\Omega_\mathrm{M}=0.27$, $\Omega_{\Lambda}=0.73$, $H_{0}=71$~km~s$^{-1}$~Mpc$^{-1}$ \citep{Komatsu_2009}, which result in a luminosity distance ($D_{L}$) and angular scale of \oj as 1.577\,Gpc and 4.48\,pc~mas$^{-1}$, respectively. 
\\


\section{Observation, Data reduction, and Imaging} \label{sec:obs_calibration}

\begin{figure*}[t]
\centering 
\includegraphics[width=\linewidth]{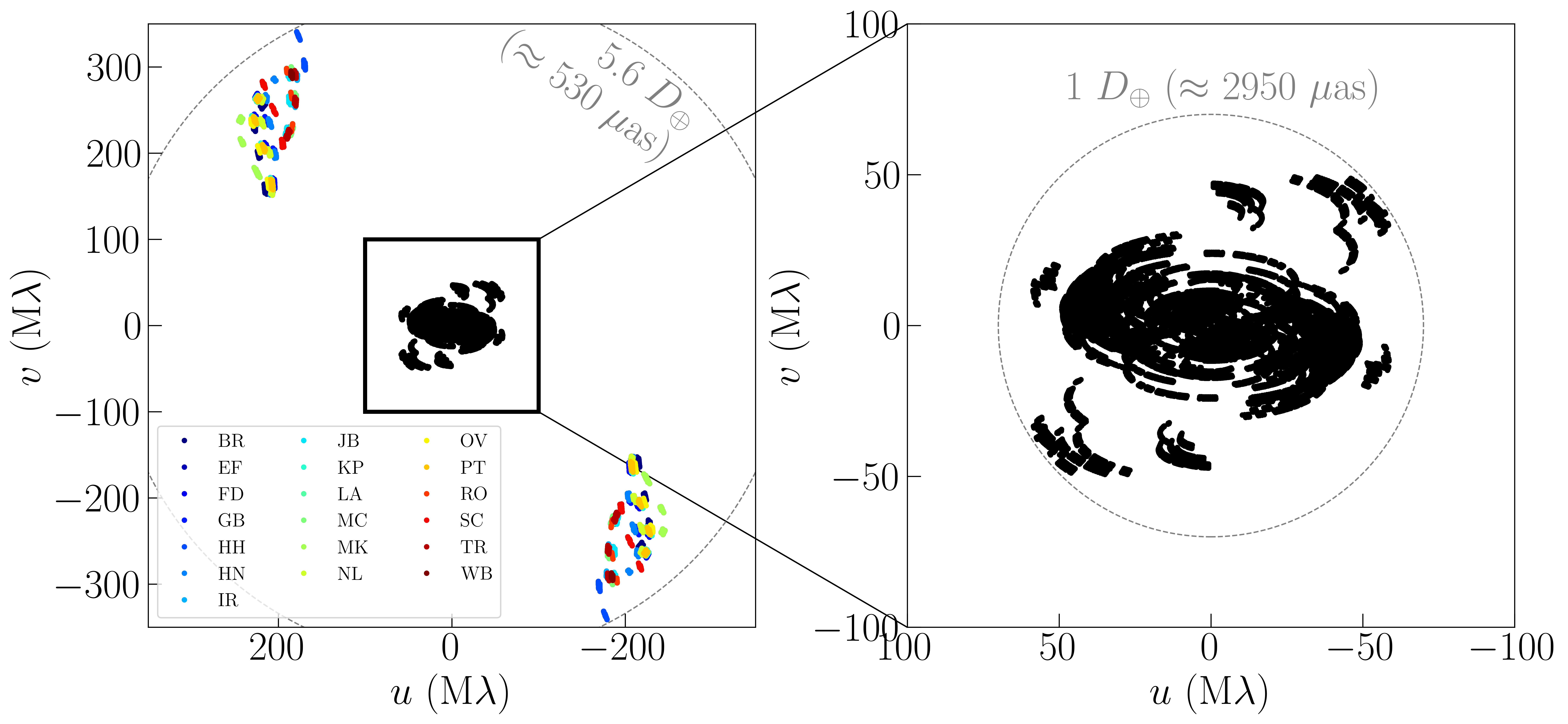}
\caption{
($u,v$)-coverage of our observation at 1.68\,GHz: full observation with ground$+$\srt (left) and ground-only array (right). Different color shows the ground stations providing the baseline with the \srt (see \autoref{tab:stations} for station codes). The maximum baseline length of space-baseline ($\sim5.6$\dearth) and ground-baseline ($\sim1$\dearth) are shown with broken line circle, which correspond to the angular resolution of $\sim530\,\mu$as and 3\,mas respectively. 
Each point has been averaged with 5\,minute interval. 
}
\label{fig:uvplt}
\end{figure*}

\subsection{Observations} \label{subsec:obs}

\oj has been observed by the \ra mission at 1.676\,GHz ($\lambda=$18.2\,cm; L-band) on 2016 April 16, from UT\,05:00 to UT\,19:00 (14\,hours in total; the experiment code GG079B), together with the calibrators 0716+714, 3C\,345, 1633+382, 3C\,84, NRAO\,150, and AO\,0235+164. The 19 telescopes on Earth and \srt have participated in the experiment (see \autoref{tab:stations}). The ground array consists of 11 telescopes in the US (10 VLBA stations and Green Bank Telescope) and 8 telescopes in Europe, so the longest Earth-based ($u,v$)-distance was $\sim$70\,M$\lambda$. 
The \srt provided the baselines longer than the Earth's diameter (\dearth) resulting in the ($u,v$)-extension up to $\sim380$\, M$\lambda$ (5.6 \dearth; see \autoref{fig:uvplt}) corresponding to an angular resolution of $\sim$0.5\,mas. The raw VLBI data were recorded with 16\,MHz per intermediate frequency (IF) band $\times$ 4 IFs \footnote{IF1: 1.620-1.636 GHz, IF2: 1.636-1.652 GHz, IF3: 1.652-1.668 GHz, IF4: 1.668-1.684 GHz} for the ground array (i.e., 64\,MHz total bandwidth) and $\times$ 2 IFs for \srt (IF3 and IF4; 32\,MHz total bandwidth), in both left- and right-handed circular polarizations (LCP and RCP, respectively). 

\begin{table}[ht]
\centering
\caption{List of stations}
\begin{tabular}{ccccc}
\hline
Name & Station & Diameter & SEFD$^{\dag}$ \\
     & Code    & (m)      & (Jy)  \\
\hline 
VLBA-Brewster & BR & 25 & 361  \\
Effelsberg & EF & 100 & 19 \\
VLBA-Fort Davis & FD & 25 & 327 \\
Green Bank Telescope & GB & 100 & 9$^{\dag\dag}$ \\
Hartebeesthoek & HH & 26 & 485 \\
VLBA-Hancock & HN & 25 & 470  \\
Irbene & IR & 32 & 1911$^{\dag\dag}$ \\
Jodrell Bank & JB & 25 & 82  \\
VLBA-Kitt Peak & KP & 25 & 489  \\
VLBA-Los Alamos & LA & 25 & 648  \\
Medicina & MC & 32 & 880 \\
VLBA-Mauna Kea & MK & 25 & 445  \\
VLBA-North Liberty & NL & 25 & 422  \\
VLBA-Owens Valley & OV & 25 & 423  \\
VLBA-Pie Town & PT & 25 & 397 \\
Spektr - R & RA$^{\ddag}$ & 10 & 2818 \\
Robledo & RO & 70 & --$^{\ddag\ddag}$  \\
VLBA-Saint Croix & SC & 25 & 318 \\
Torun & TR & 32 & 760 \\
Westerbork & WB & 25 & 443 \\
\hline
\end{tabular}
\\ \vspace{0.3cm}
\raggedright{
$\dag$ The values are calculated from the averaged system temperature, characteristic gain curve and the effective aperture size. \\
$\ddag$ In the data, this is labeled as R1 and R2 for each of the two Earth-based sub-arrays. \\
$\dag\dag$ These are estimated from RCP-only (see \autoref{subsec:calibration}), while the others are averaged across dual polarization. \\
$\ddag\ddag$ No system temperature measurements. 
}
\label{tab:stations}
\end{table}

\subsection{Data reduction} \label{subsec:calibration}

The observational data consist of two sets corresponding to two Earth-based sub-arrays, the EVN and VLBA. Both arrays consistently observed \oj, while the calibrators were observed intermittently by each sub-array. The data were first processed using the {\tt DiFX} software correlator with a dedicated version for \ra \citep{Bruni_2016}. 

For the post-correlation data reduction, the NRAO Astronomical Imaging Processing System (AIPS; \citealt{Greisen_2003}) has been used. First, the cross-correlated amplitudes were corrected using the auto-correlated amplitudes ({\tt ACSCL}) and the bandpass for only amplitudes were corrected ({\tt BPASS}). Note that the auto-correlation amplitudes at this step have a slight offset from unity so a small adjustment needs to be applied with {\tt ACSCL}\footnote{VLBA Scientific Memo 37}. For the a-priori amplitude calibration, we used the {\tt AIPS APCAL task} with available system temperature and antenna gain information following the calibration method described in  \citealt{Cho_2017}. The atmospheric opacity correction was not applied since the effect is negligible at the frequency of 1.7\,GHz. At this step we found that the amplitudes at IF1 showed larger scatter at all stations. The measured system temperatures ($T_{\rm sys}$) were also much more scattered than in the other IFs, and the task could not correct the amplitudes properly (see \autoref{app:error}). A similar behavior was found in the data from JB (all IFs) and a few VLBA stations (KP, LA, MC in IF4). This station-based problem has been treated later at the self-calibration stage of the processing using the reconstructed image. 

As for the phase calibration, the ionospheric effect (only for the ground array) and parallactic angle corrections ({\tt VLBATECR} and {\tt VLBAPANG}, respectively) were first applied. The instrumental phase and delay offset correction was made using the best fringe fit solutions on a short segment of data ({\tt FRING}). The global fringe search was done with the ground array first, and then further searched for the ground-space baselines. Due to the different acceleration of \srt in its orbit, different solution intervals were applied for each scan (e.g., 10$-$180 seconds). As a result, the phase delay and fringe frequency solutions were successfully obtained with signal-to-noise ratio (S/N) $>$\,5  up to a baseline length of $\sim385\,{\rm M}\lambda$. With this, the bandpass calibration was done for both amplitudes and phases using the cross-power spectra. The delay between LCP and RCP was obtained through {\tt RLDLY}. 

\begin{figure}[t]
\centering 
\includegraphics[width=\columnwidth]{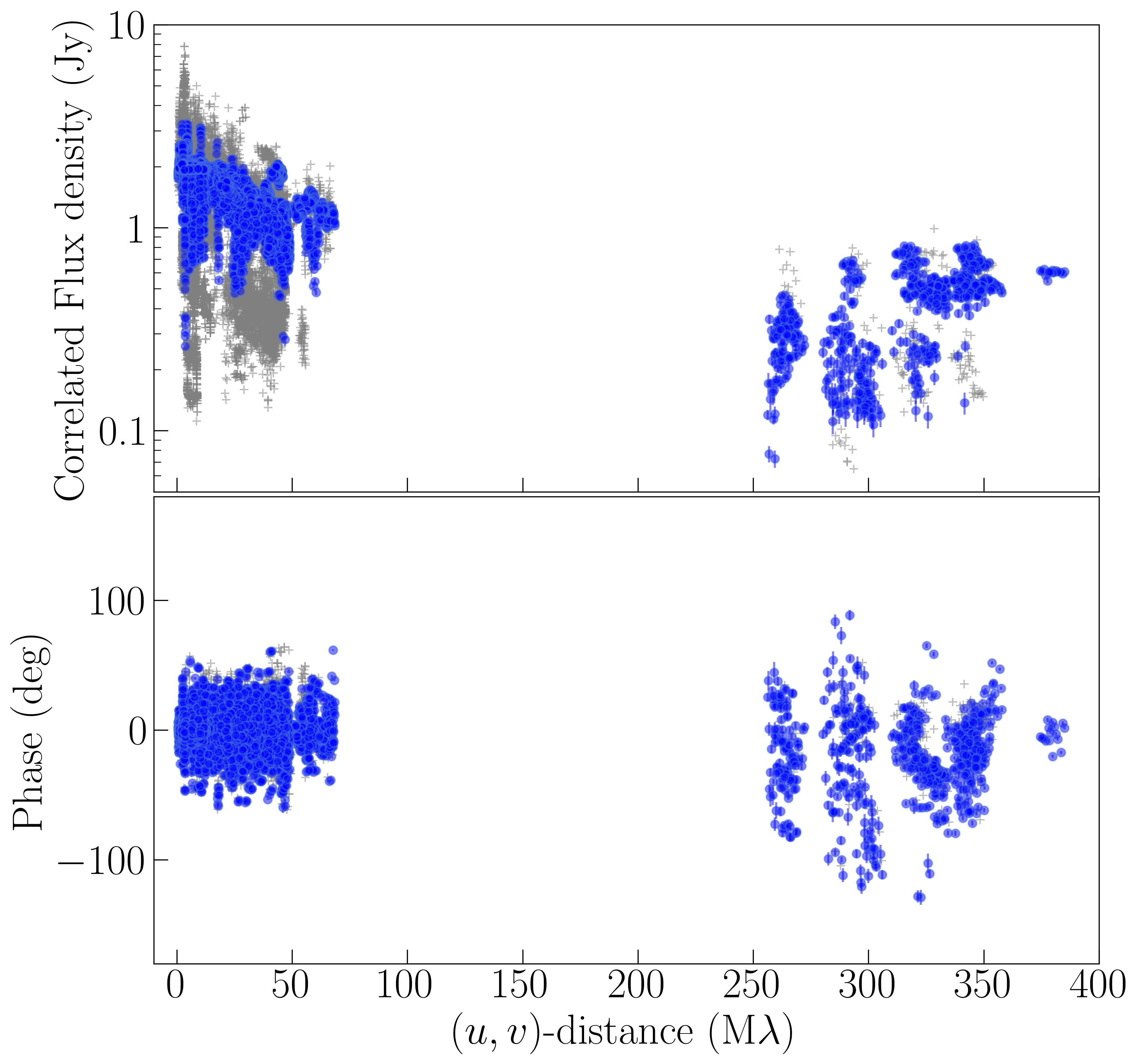}
\caption{
The visibility amplitudes (upper) and phases (lower), as a function of the ($u,v$)-distance. 
Each point has been averaged with 5\,minute interval. Note that only the blue, circle data points are used for CLEAN imaging, and all data including the gray, cross points are self-calibrated with the image (see \autoref{sec:obs_calibration}). 
}
\label{fig:radpl}
\end{figure}

After the calibrations, the data were averaged within each IF and coherently time averaged over 5\,minutes (\autoref{fig:radpl} for the visibility amplitudes and phases). Before imaging, outliers and low S/N points (e.g., S/N $<$ 7) were flagged out. The RO station recorded single-polarization at each IF -- in particular in the format of RCP-LCP-LCP-RCP at IF1$-$4, espectively. However, a phase-offset issue was identified at IF1 and IF4, resulting in the utilization of only IF2 and IF3 data with LCP polarization. At the GB station, the L-band receiver encountered an issue that caused a phase difference of $\sim90^\circ$ in each polarization, resulting in near-zero signal for the LCP polarization. As a result, only RCP data were utilized as pseudo Stokes I for imaging. A similar issue was also identified at the IR station, and only RCP data were used accordingly. 

\subsection{Imaging} \label{subsec:imaging}

For imaging, we have applied a conventional VLBI imaging software, \difmap \citep{Shepherd_1994}, which is based on the classical CLEAN algorithm. This models the image as a collection of point sources from the inverse Fourier transform of the sampled visibilities. 

We first implemented the iterative CLEAN and self-calibration with the ground array. Note that, at this step, the data showing spurious amplitudes (i.e., IF1 at all stations, all IFs at JB, and IF4 at KP, LA, MC) and having receiver issues (i.e., RO, GB, IR) were first flagged out. Then the CLEAN imaging was implemented with the remaining data. After that, full data including the flagged points (i.e., unflag) were self-calibrated with the obtained image. After the phase and amplitude self-calibration for all ground stations with the solution interval down to 5\,minutes, the CLEAN models were removed and the \srt was included for imaging so that we could have the higher angular resolution. In this step, the same CLEAN windows from ground array imaging were used and only the \srt was self-calibrated for both phase and amplitude. This prevents the over-substraction of the CLEAN components. We have tested different $uv$-weighting parameters for both uniform ({\tt UVWEIGHT}=[2,0], [2,-1] [2,-2]) and natural ({\tt UVWEIGHT}=[0,-1], [0,-2]) weighting, and found the best parameter to be {\tt UVWEIGHT}=[2,-2], providing the lowest $\chi_{\nu}^{2}$ for closure quantities. Note however that the $\chi_{\nu}^{2}$ differences are small (e.g., $<0.1$) between different choices of the power of {\tt UVWEIGHT}, while relatively larger difference is found between uniform and natural weighting. 
\\


\section{Results} \label{sec:results}

\subsection{Structure of \oj}\label{sec:Iim}

\subsubsection{Stokes~I image}

\begin{figure*}[t]
\centering 
\includegraphics[width=\linewidth]{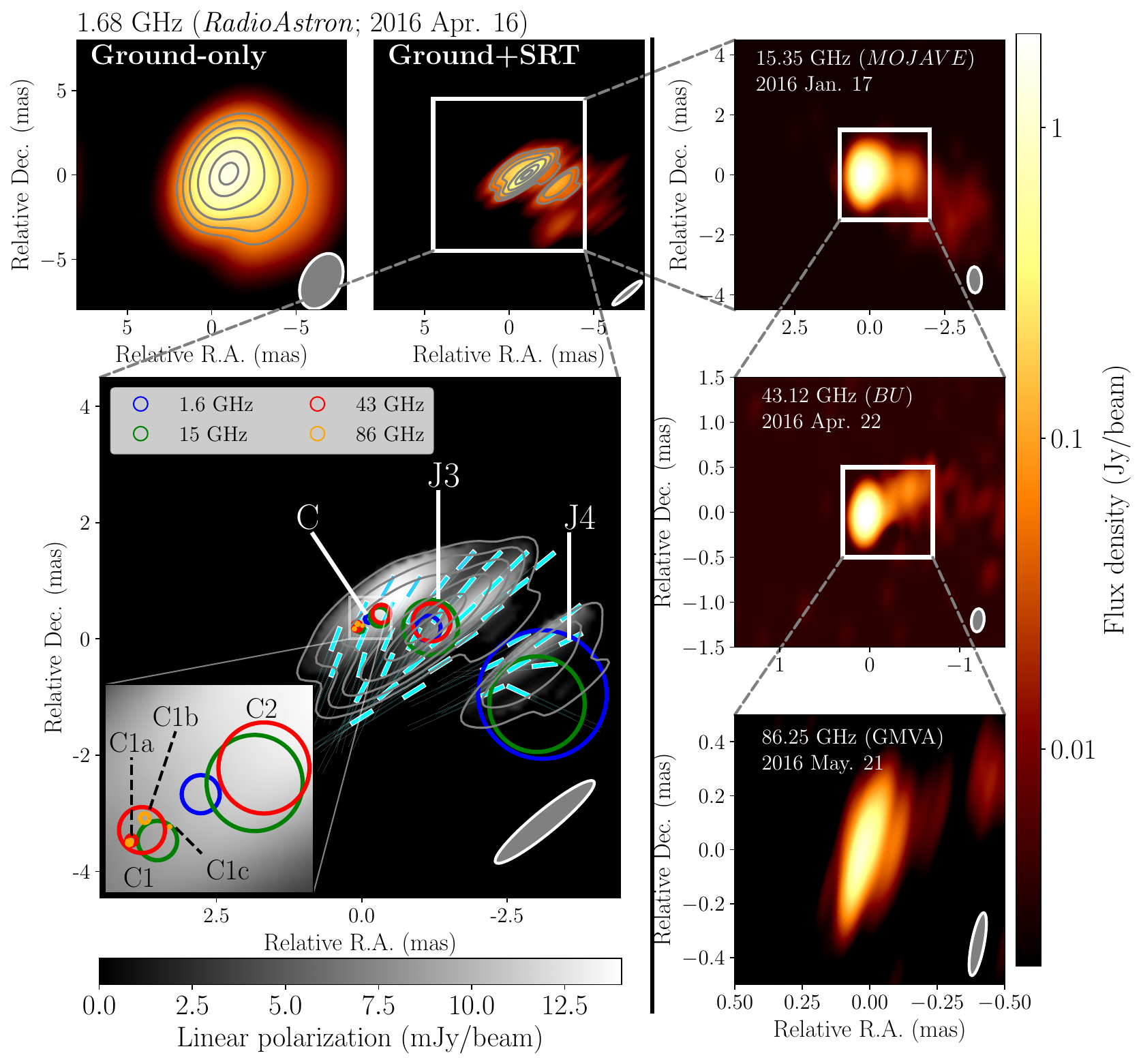}
\caption{
(Left) Stokes I images of \oj at 1.68\,GHz with a nominal resolution of ground-array only and ground$+$\srt (top). Bottom panel shows the ground$+$\srt image with total intensity (contour), model fitting results (colored circles), linear polarization (gray color), and EVPA (ticks). Note that the total intensity contours are set to 4, 8, 16, 32, 64, 90\,\% of the peak intensity. Reliable linear polarization are obtained at where the total intensity is larger than 10\,$\sigma$ of the root-mean-squared flux density of residual image. 
(Right) The higher frequency images are shown for comparison: 15.35\,GHz image from MOJAVE project (top), 43.12\,GHz image from VLBA-BU-BLAZAR project (middle), and 86.25\,GHz image from \citet{Lico_2022}. The field-of-view of image gets smaller at higher frequency which corresponds to the white box in the panels of lower frequency. The restoring beam is shown at the bottom-right corner of each panel. The color scale is in Jy per beam for total intensity images, and in milli-Jy per beam for linear polarization. 
}
\label{fig:image_I}
\end{figure*}

\autoref{fig:image_I} (left) shows the Stokes~I image of \oj at 1.68\,GHz from our observation. Two images at top show the resultant image with the ground-array only (3.5$\times$2.3\,mas, PA$=-$27\fdg3) and ground$+$\srt from uniform-weighted nominal beam (2.2$\times$0.4\,mas, PA$=-$50\fdg5). The left, bottom figure shows the ground$+$\srt image with total intensity (contour), model fitting results (colored circles; see \autoref{sec:modelfit}), linear polarization (gray color; see \autoref{sec:polim}), and electric vector position angle (EVPA). The adjacent higher frequency images are also presented in \autoref{fig:image_I} (right): (from top to bottom) the 15.35\,GHz image in 2016 Jan.~17 from the MOJAVE project\footnote{https://www.cv.nrao.edu/MOJAVE/} \citep{Lister_2018}, 43.12\,GHz image in 2016 Apr.~22 from the VLBA-BU-BLAZAR monitoring project\footnote{https://www.bu.edu/blazars/} \citep{Jorstad_2017, Weaver_2022}, and 86.25\,GHz image in 2016 May~21 from the GMVA monitoring program\footnote{https://www.bu.edu/blazars/vlbi3mm} with the targets of the VLBA-BU-BLAZAR \citep{Lico_2022}. 

In our \ra image at 1.68\,GHz, three main emission features are found: 
1) one component at the center, 2) the brightest component at the west, and 3) a distinct component toward south-west at $\sim1.6\,$mas. Due to the elongated ($u,v$)-coverage (\autoref{fig:uvplt}), we have a better resolution in the northeast-southwest direction than to northwest-southeast (i.e., PA of nominal beam $=-$50\fdg5), which enables us to well resolve the features. The image fidelity, with results from fitting to closure quantities and gain corrections for each telescope, is discussed in \autoref{app:error}. Comparing with the images at higher frequencies and considering the bent-jet structure of \oj \citep[e.g.,][]{Gomez_2022}, the center component is thought to be the core region, while the brightest component at the west is an optically thin jet emission. Note that our identification of the radio core is supported by its compact structure (usually at the most upstream of a jet), a flat spectrum (\autoref{sec:spix}), and weak polarization (\autoref{sec:polim}). 

\begin{figure*}[t]
\centering 
\includegraphics[width=\linewidth]{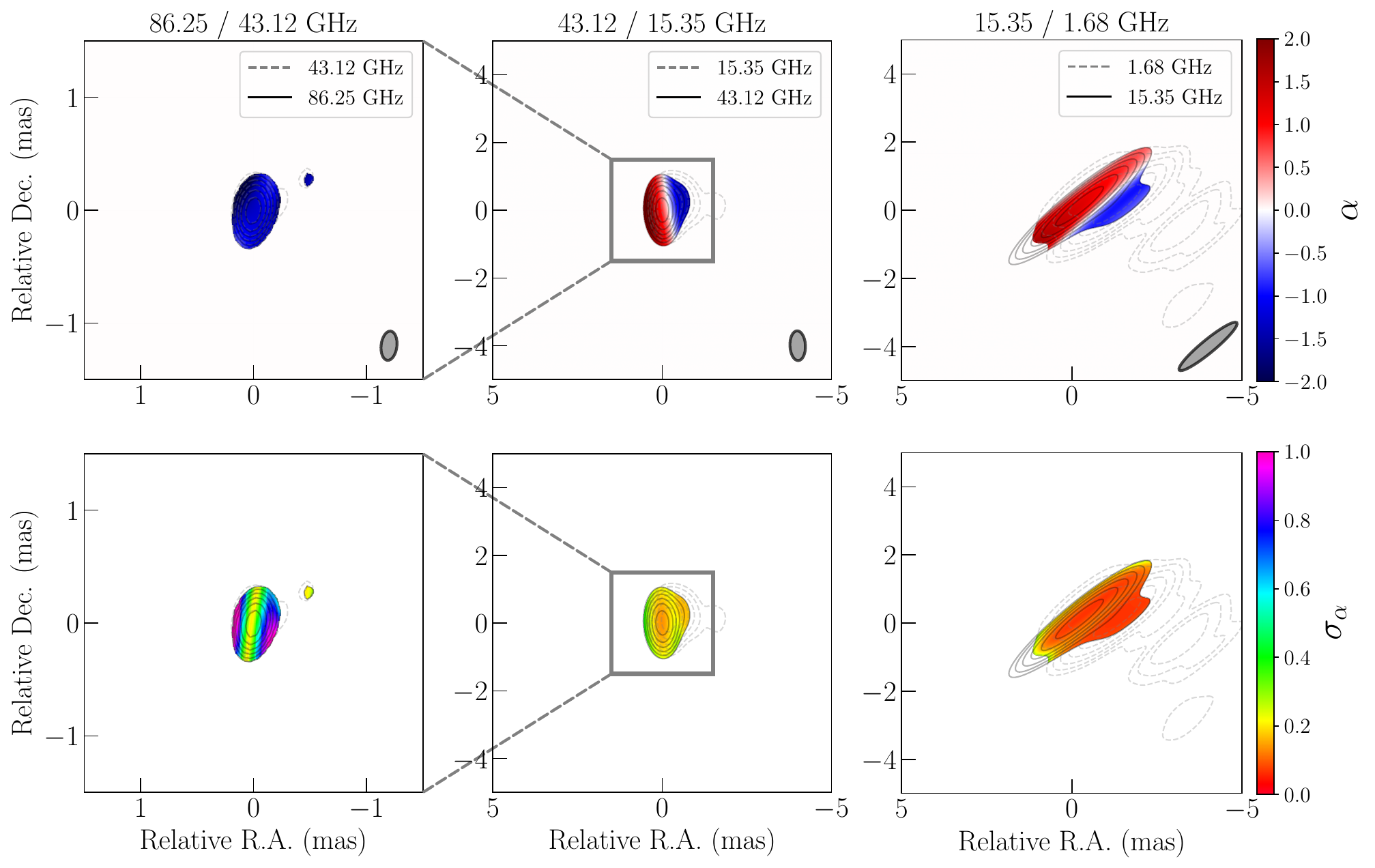} 
\caption{
Spectral index map (top) and corresponding error map (bottom) at three frequencies pairs: 43.12$-$86.25\,GHz (left), 15.35$-$43.12\,GHz (middle), and 1.68$-$15.35\,GHz (right). The color scale indicates the spectral index, $\alpha$, and its error, $\sigma_\alpha$, as shown at right of top and bottom panels, respectively. The contours of the source structure at each frequency are shown in each panel with gray, broken line for the lower frequency and black, solid line for the higher one. The contours are set to 2, 4, 8, 16, 32, 64\,\% of the peak intensity. Note that the relative position of each image has been aligned according to cross correlation of core-masked images, as described in \autoref{sec:Iim}. 
}
\label{fig:spix_map}
\end{figure*}

\subsubsection{Spectral index} 
\label{sec:spix}

To confirm the distribution of optical depth, we have obtained the spectral index map by aligning images at two different frequencies. The alignment has been achieved using two-dimensional cross correlation of the optically thin extended jet structure \citep{Walker_2000, Croke_2008}. Since the ($u,v$)-spacing and the nominal angular resolutions are different in the original images at two different frequencies, the minimum and maximum ($u,v$)-distances were first matched with each other to be the same (e.g., {\tt UVRANGE} in \difmap; e.g., \citealt{Fromm_2013}). Next, two images at different frequencies have been convolved with a nominal beam corresponding to the angular resolution at a lower frequency and presented with the same pixel scale ($\lesssim1/20$ of the lower frequency beam diameter; e.g., \citealt{Kim_2014}). After this, the two images have been cross correlated and the higher frequency image has been shifted to achieve the maximum cross correlation. Note that a large field-of-view (FoV) is used so that entire image structure can be shifted within the FoV. Before calculating the cross correlation, the core region which is thought to be partially optically thick \citep[e.g.,][]{Pushkarev_2012} has been masked so that the cross correlation is obtained only for the optically thin structures. To limit the effect of residual structures, the flux density cutoff has been applied at the level of 10\,$\sigma$ of the root-mean-squared flux density of residual image. 

By finding all relative shifts, the images are aligned with each other and the spectral index map is determined (\autoref{fig:spix_map}, top). This reveals that the center component at 1.68\,GHz (e.g., component C in model-fitting; see \autoref{fig:image_I} and \autoref{sec:modelfit}) corresponds to the optically thick core region with spectral index $\alpha\gtrsim1$, in $S_{\nu}\propto\nu^{\alpha}$ where $S_{\nu}$ and $\nu$ are flux density and observing frequency respectively (i.e., an inverted spectrum). This inversion extends up to 43\,GHz, and explains the faint nature of the core region at 1.68\,GHz due to the opacity effect. The brightest west and the south-west components (i.e., components J3 and J4 in model-fitting) correspond to optically thin jet features with a steep spectrum. Two major uncertainties are considered in the estimation of the spectral index: a systematic error from image alignment and a random error from the image noise \citep[e.g.,][]{Pushkarev_2012b,Fromm_2013,Hovatta_2014}. The former error is mainly affected by the size of core mask. Note that, to obtain a reasonable cross-correlation, the core region needs to be completely masked. For this, 1.5\,times the convolution beam size is found as the minimum size of the mask. To estimate the effect of the mask size, the relative image alignments and corresponding spectral index maps are obtained using different sizes of the mask ranging from 1.5 to 2.5 times the convolution beam size. The standard deviation of resultant images are then considered as the systematic error. To account for an additional alignment error, the image providing the highest cross-correlation is manually shifted within 1/3 of beam size, and the standard deviation of the spectral index is further obtained. On the other hand, the random error is estimated from the image noise and flux density measurement error \citep{Lico_2012}. As a result, the total error of the spectral index map is obtained by the quadratic sum of these uncertainties (\autoref{fig:spix_map}, bottom).

\begin{table*}[ht]
\centering
\addtolength{\tabcolsep}{-3pt}
\small
\caption{Model fitting results} 
\begin{tabular}{ccccccccccc}
\hline
\hline
Component & Freq. &  Sub-Comp.1 &  Sub-Comp.2 &            S &           r &         PA &        FWHM &     \tb  &  $m$  &  EVPA  \\
  & (GHz) &  &  & (mJy) & ($\mu$as) & ($^\circ$) & ($\mu$as) & (10$^{10}$ K) & (\%) & ($^\circ$) \\
(1) & (2) & (3) & (4) & (5) & (6) & (7) & (8) & (9) & (10) & (11) \\
\midrule
       C &    1.6 & ... & ... &   440$\,\pm\,$90 &   280$\,\pm\,$10 &  $-$54$\,\pm\,$1 &         $<\,$131 &         $>\,$1413 & 0.1$\,\pm\,$0.1 & $-$24$\,\pm\,$11  \\
       ... &   15.0 & C1 & ... &  2530$\,\pm\,$270 &     50$\,\pm\,$1 &  $-$90$\,\pm\,$1 &    136$\,\pm\,$2 &     92$\,\pm\,$6 & ... & ...  \\
       ... &   15.0 & C2 & ... &   860$\,\pm\,$160 &   430$\,\pm\,$1 &  $-$62$\,\pm\,$1 &   332$\,\pm\,$7 &       5$\,\pm\,$1 & ... & ...  \\
       ... &   43.0 & C1  & C1a & 3080$\,\pm\,$530 &      ... &  ... &          $<\,$35 &          $>\,$205 & ... & ...  \\
       ... &   43.0 & ... & C1b &  1120$\,\pm\,$320 &    50$\,\pm\,$1 & $-$45$\,\pm\,$3 &   158$\,\pm\,$6 &       4$\,\pm\,$1 & ... & ...  \\
       ... &   43.0 & C2 & ... &   260$\,\pm\,$170 &   520$\,\pm\,$30 &  $-$61$\,\pm\,$2 &  314$\,\pm\,$53 & 0.2$\,\pm\,$0.1 & ... & ...  \\
       ... & 86.0 & C1 & C1a & 1080$\,\pm\,$220 & ... &  ... & 17 & 80 & ... & ...  \\
       ... &  86.0 & \citep{Lico_2022} & C1b & 530$\,\pm\,$110 & 100$\,\pm\,$10 & $-$32 & 33 & 10 & ... & ...  \\
       ... & 86.0 & ... & C1c & 60$\,\pm\,$10 & 150$\,\pm\,$10 & $-$68 & 4 & 80 & ... & ...  \\
       J3 &    1.6 & ... & ... &  1120$\,\pm\,$140 &  1240$\,\pm\,$1 &  $-$90$\,\pm\,$1 &   435$\,\pm\,$4 &   334$\,\pm\,$40 & 0.3$\,\pm\,$0.1 & $-$33$\,\pm\,$44  \\
       ... &   15.0 & ... & ... & 240$\,\pm\,$90 &  1260$\,\pm\,$30 &  $-$89$\,\pm\,$1 &  944$\,\pm\,$54 &  0.2$\,\pm\,$0.1 & ... & ...  \\
       ... &   43.0 & ... & ... &    80$\,\pm\,$80 & 1320$\,\pm\,$190 & $-$85$\,\pm\,$8 & 655$\,\pm\,$387 & 0.02$\,\pm\,$0.01 & ... & ...  \\
       J4 &    1.6 & ... & ... &   420$\,\pm\,$80 &  3410$\,\pm\,$20 & $-$109$\,\pm\,$1 & 2212$\,\pm\,$43 &       5$\,\pm\,$1 & 2.9$\,\pm\,$1.2 & $-$70$\,\pm\,$60  \\
       ... &   15.0 & ... & ... &    80$\,\pm\,$50 & 3350$\,\pm\,$140 & $-$112$\,\pm\,$2 & 1644$\,\pm\,$273 & 0.02$\,\pm\,$0.01 & ... & ...  \\
\hline
\end{tabular}
\\ \vspace{0.3cm}
\raggedright{
\scriptsize{
\textbf{Note. } From left to right columns: 
(1) Component ID, 
(2) Observing frequency in GHz, 
(3-4) Sub-component ID, 
(5) Flux density in mJy, 
(6) Radial distance from the core (i.e., C1a) in $\mu$as, 
(7) Position angle of the component position relative to the core in degree, 
(8) Component size in $\mu$as, 
(9) Brightness temperature in source rest frame in K, 
(10) Fractional linear polarization in \%, and 
(11) Electric vector position angle in degree. 
}}
\label{tab:modelfit}
\end{table*}

\begin{table*}[ht]
\centering
\caption{Spectral analysis of model fitting components}
\begin{tabular}{cccccccc}
\hline
\hline
Comp. 
& $\alpha_{1.6-15\,{\rm GHz}}$ 
& $\alpha_{15-43\,{\rm GHz}}$ 
& $\alpha_{43-86\,{\rm GHz}}$ 
& S$_{m}$ 
& $\nu_{m}$ 
& $B_\mathrm{ssa}$
& $B_\mathrm{eq}$
\\
&&&
& [Jy] 
& [GHz] 
& [G] 
& [G] 
\\
(1) & (2) & (3) & (4) & (5) & (6) & (7) & (8) \\
\hline 
C & 0.9$\,\pm\,$0.1 & 0.3$\,\pm\,$0.2 & $\geqslant-1.4$ & $\geqslant\,$4.4 & $\geqslant\,$28 & ... & ... \\
C1 & $>0.8$ & 0.5$\,\pm\,$0.3 & $-$1.3$\,\pm\,$0.4 & 4.0$\,\pm\,$0.6 & 33.0$\,\pm\,$1.0 & $0.3\,\pm\,0.1\,\delta_\mathrm{j}$ & 
$35.6\,\pm\,1.8\,\delta_\mathrm{j}^{\,-8/7}$ \\ 
C2 & $>0.3$ & $-$1.2$^{+0.7}_{-1.1}$ & ... & 0.7$\,\pm\,$0.2 & 11.5$\,\pm\,$1.5 & $0.1\,\pm\,0.08\,\delta_\mathrm{j}$ & 
$22.8\,\pm\,2.9\,\delta_\mathrm{j}^{\,-8/7}$ \\
J3 & $-$0.7$^{+0.2}_{-0.3}$ & $-$1.1$^{+1.3}_{-4.5}$ & ... & $\geqslant1.2$ & $\leqslant0.8$ & ... & ... \\
J4 & $-$0.7$^{+0.3}_{-0.5}$ & ... & ... & ... & ... & ... & ... \\ 
\hline
\end{tabular}
\\ \vspace{0.3cm}
\raggedright{
\scriptsize{
\textbf{Note. } From left to right columns: 
(1) Component ID, 
(2) Spectral index between 1.6 and 15\,GHz, 
(3) Spectral index between 15 and 43\,GHz, 
(4) Spectral index between 43 and 86\,GHz, 
(5) Turnover flux density in Jy, 
(6) Turnover frequency in GHz, 
(7) Magnetic field strength from SSA in G, and 
(8) Magnetic field strength from equipartition in G. 
$\delta_\mathrm{j}$ is a doppler factor. 
}}
\label{tab:spectral}
\end{table*}

\begin{figure*}[ht]
\centering 
\includegraphics[width=\linewidth]{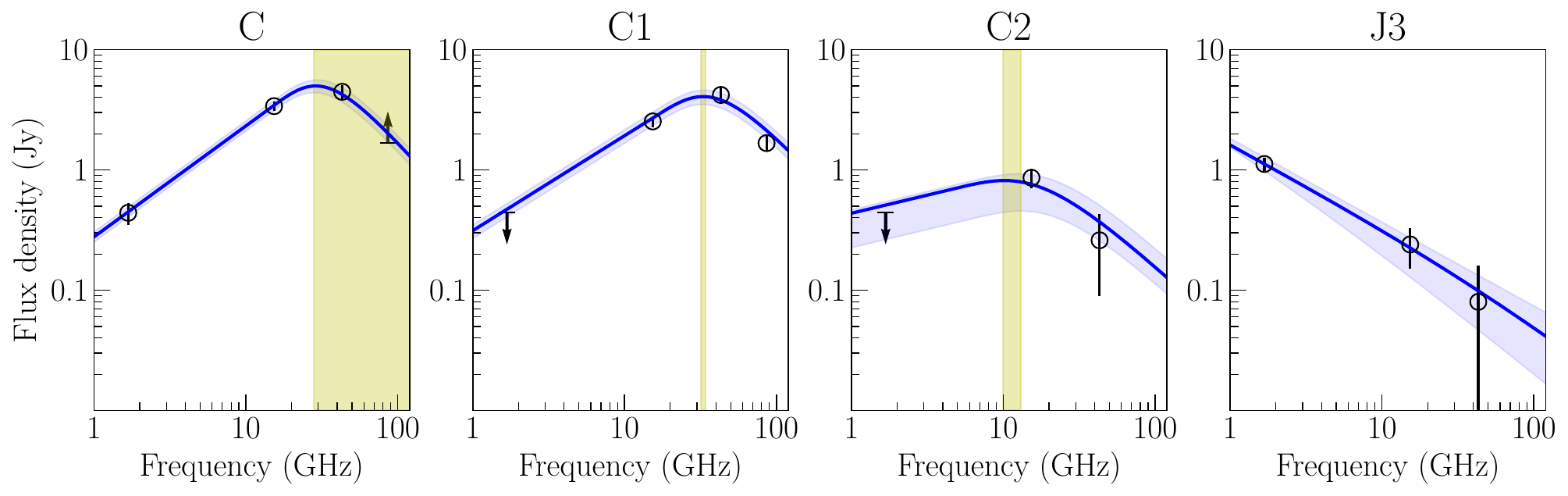}
\caption{
Spectra of model fitting components identified at 1.68, 15.35, 43.12, and 86.25\,GHz: (from left to right) C, C1, C2, and J3. Flux densities are shown with black, circle points with uncertainty (see \autoref{tab:modelfit}), and the blue, solid-line shows the fitting result assuming a SSA spectrum (see \autoref{sec:modelfit}). Fitting errors, due to the flux density uncertainties, are presented with the blue, shaded area. The derived SSA turnover frequencies in C, C1, and C2 are shown with the yellow, shaded area. 
} 
\label{fig:spix_comp}
\end{figure*}

\subsubsection{Model fitting} 
\label{sec:modelfit}

The images at each frequency were modeled with {\tt modelfit} procedure in \difmap (\autoref{tab:modelfit}), except the 86.25\,GHz where the model fitting results were already provided by \citet{Lico_2022}. With the self-calibrated data from a reconstructed image, the circular Gaussian models were fitted to the complex visibilities using Levenberg-Marquardt non-linear least squares minimization until the residual peak intensity reached less than 5\,times the root-mean-squared flux density in residual image. In this way, at 1.68\,GHz, we found that three circular Gaussian components, labeled as C, J3 and J4, describe the source's morphology effectively. The $\chi_{\nu}^{2}$ of closure phases for the case with three Gaussian components converges to $\sim0.8$ from $\sim25$ for the one Gaussian component case. Note that it was difficult to find more model components beyond three Gaussians, as the initial guess (e.g., based on the residual peak intensity) became uncertain and the result easily converged to negative flux density. 

Component C is resolved at 15.35 and 43.12\,GHz into two sub-components, C1 and C2. In addition, the C1 component is further resolved into two sub-components (C1a and C1b) at 43.12\,GHz and three sub-components (C1a, C1b, and C1c) at 86.25\,GHz (\autoref{fig:image_I}). That is, the component C is blended with multiple sub-components and has positive spectral index supporting the identification as a core region with optically thick {spectrum} up to 43.12\,GHz. On the other hand, the components J3 and J4 have steep spectra indicating that they are optically thin jet components. Model fitting uncertainties shown in columns 5--11 of \autoref{tab:modelfit} are estimated following the method described by \citet{Schinzel_2012}. \autoref{tab:spectral} summarizes spectral indices of the model components which appear to be consistent with the spectral index distributions shown in \autoref{fig:spix_map}. 

In \autoref{fig:image_I} (bottom, left), the model components at four frequencies are shown with different colors. Note that the relative position of model components across frequencies are aligned based on the image alignment (see \autoref{sec:spix}). With this, the position of the C1a component is well consistent at 43 and 86\,GHz (e.g., $\lesssim10\,\mu$as) that can be identified as the innermost core. The distances of the other components are then estimated relative to the C1a (column~6 in \autoref{tab:modelfit}), which allow us to estimate the frequency dependent radio core position shift, the so called core shift \citep[e.g.,][]{Lobanov_1998}. At 15\,GHz, the core corresponds to the most upstream component C1 so the core shift relative to C1a is obtained as $\sim50\,\mu$as. At 1.68\,GHz, on the other hand, only the blended core component C is found, so the core shifts from 15 and 43 (or 86)\,GHz are obtained as $\sim230$ and $\sim280\,\mu$as, respectively. 

By fitting the synchrotron self-absorbed (SSA) spectrum with a power-law energy distribution \citep[e.g.,][]{Turler_2000, Gomez_2022}, the maximum synchrotron turnover flux density in Jy, $S_\mathrm{m}$, and turnover frequency in GHz, $\nu_\mathrm{m}$, are found for each component (\autoref{tab:spectral}): 

\begin{equation}
 S_{\nu} = S_\mathrm{m} \left( \dfrac{\nu}{\nu_\mathrm{m}} \right)^{\alpha_\mathrm{thick}}
 \dfrac{ 1 - \exp[-\tau_\mathrm{m}(\nu / \nu_\mathrm{m})^{\alpha_\mathrm{thin} - \alpha_\mathrm{thick}}]}{1 - \exp (-\tau_\mathrm{m}) } ~~~~ [\mathrm{Jy}],
 \label{eq:Sm}
\end{equation}
where $S_{\nu}$ is the observed flux density in Jy, $\alpha_\mathrm{thick}$ and $\alpha_\mathrm{thin}$ are the spectral indices of the optically thick and thin parts of the spectrum, respectively. $\tau_\mathrm{m}$ is the optical depth at $\nu_\mathrm{m}$, which is defined as $\sim 3/2 \left[ \left( 1-\left(8 \alpha_\mathrm{thin} / 3\alpha_\mathrm{thick}\right) \right)^{1/2} - 1 \right]$. Note that $S_\mathrm{m}$ is defined as $S_\mathrm{m}^\mathrm{thin}\,(1 - \mathrm{exp}(-\tau_\mathrm{m}))\,/\,\tau_\mathrm{m}$, where $S_\mathrm{m}^\mathrm{thin}$ is the extrapolated flux density of optically thin spectrum at $\nu_\mathrm{m}$ \citep[e.g.,][]{Turler_2000}.  

\begin{figure}[t]
\centering
\includegraphics[width=\columnwidth]{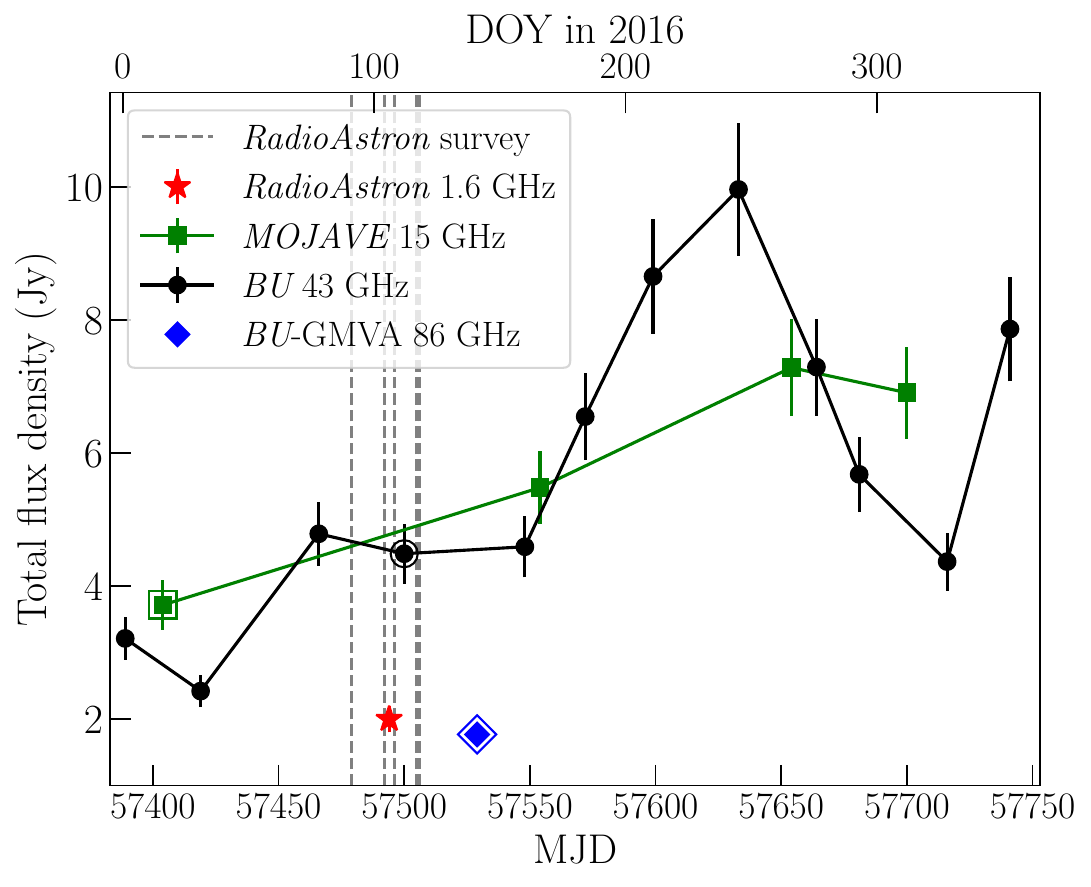}
\caption{
\oj light curve in 2016 from the VLBA observations at 15\,GHz (MOJAVE; green square, solid line) and 43\,GHz (VLBA-BU-BLAZAR; black circle, solid line). Our \ra observation at 1.68\,GHz is shown with red asterisk, and the GMVA 86\,GHz observation is shown with blue diamond. The observations compared in this study are marked with an additional edge. The vertical broken lines show the epochs of adjacent \ra snapshot sessions. 
}
\label{fig:light_curve}
\end{figure}

First, the flux densities of the component C at 15, 43, 86\,GHz are assumed as sum of flux densities of sub-components. Since the C2 component is not found at 86\,GHz, however, the flux density at 86\,GHz is considered as lower limit. With this, the component C shows a transition from optically thick to thin at $\nu_{m}\geqslant28\,$GHz with $S_\mathrm{m}\geqslant4.4\,$Jy (see \autoref{tab:spectral} and \autoref{fig:spix_comp}). 
As for the sub-components in core region, C1 and C2, the spectral index with 1.68\,GHz is obtained assuming the flux density of C as an upper limit. Then the opacity transition is found in the component C1 at $\nu_{m}\approx33\,$GHz with $S_\mathrm{m}\approx4.0\,$Jy, and in the component C2 at $\nu_{m}\approx11.5\,$GHz with $S_\mathrm{m}\approx0.7\,$Jy. As for the J3 component, it is optically thin across 1.68$\,-\,$43\,GHz so the lower and upper limit of S$_{m}\gtrsim1.2\,$Jy and $\nu_{m}\lesssim0.8\,$GHz are found, respectively. Note that the optically thick spectral index, $\alpha_{\rm t}$, in the SSA spectrum for J3 is assumed to be the same as for C (i.e., 0.3). 
It is also worth noting that the small number of data points and differences in observing dates (up to $\sim3\,$months between 1.6 and 15\,GHz) may introduce further uncertainties in the estimates. According to the light curve at 43\,GHz (\autoref{fig:light_curve}), for instance, the flux variation is stable during Apr$-$May~2016 so the uncertainties arising from time gap might be relatively small at 1.68, 43, and 86\,GHz. On the other hand, large flux decrease is shown on Jan.~2016 so the uncertainty of 15\,GHz may become larger up to $\sim30\,\%$.

\subsection{Brightness temperature} \label{subsec:tb}

\begin{figure}[t]
\centering
\includegraphics[width=\columnwidth]{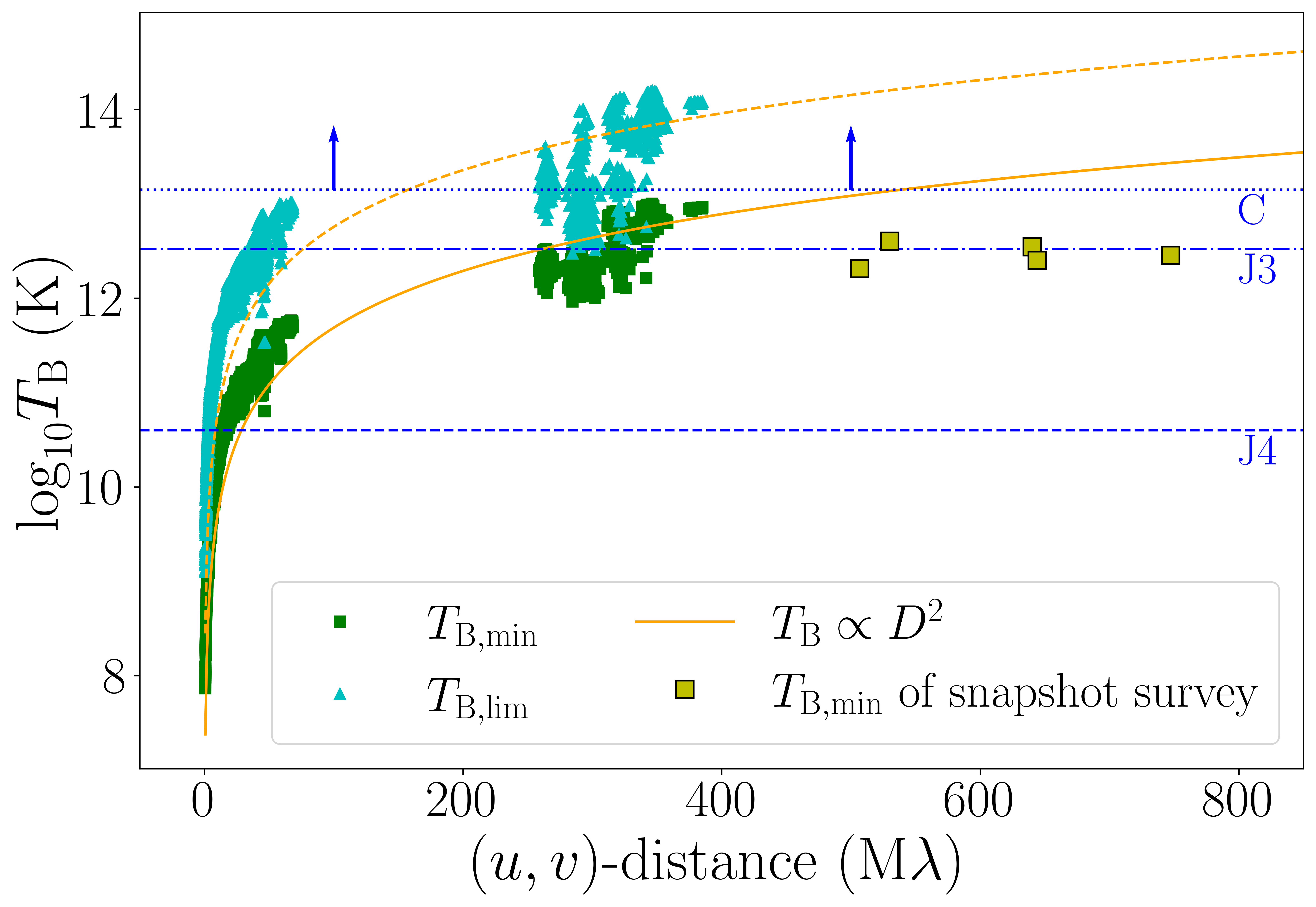} 
\caption{The brightness temperature as a function of the ($u,v$)-distance, $D$. The green, square shows the minimum \tb and the cyan, triangle shows the limiting \tb. The yellow, square shows the $T_{\rm B,min}$ from five \ra snapshot survey sessions close to our observation. 
The blue, horizontal lines show the \tb at model components C (dotted), J3 (dotted-broken), and J4 (broken). Note that a lower limit is reported for C, since its size is smaller than the resolution limit (see \autoref{subsec:tb}). 
}
\label{fig:btemp}
\end{figure}

The brightness temperature of each component in the rest frame of the source is then estimated from the fitted flux densities and sizes \citep[e.g.,][]{Gomez_2022}, 
\begin{equation}
    T_{\rm B} = 1.22\times10^{12} \frac{S}{\theta_{\rm obs}^{2}\nu^{2}} (1+z) \hspace{0.3cm} [\rm K]
\end{equation}
where $S$ is the component flux density in Jy, $\theta_{\rm obs}$ is the full-width-half-maximum (FWHM) size of the model component in mas, $\nu$ is the observing frequency in GHz, and $z$ is the source redshift. Note that the resolution limit has been used for the size of unresolved model components \citep{Lobanov_2005} so that the lower limit of $T_{\rm B}$ is estimated for such cases (see column~8 and 9 of \autoref{tab:modelfit}). Additionally, the minimum and limiting brightness temperatures ($T_{\rm B,min}$ and $T_{\rm B,lim}$, respectively) are estimated directly from the visibility amplitudes \citep{Lobanov_2015} so that a range of actual \tb can be constrained (\autoref{fig:btemp}). 

In addition to imaging observations, \oj was regularly monitored by \ra with the visibility tracking mode as part of the AGN survey program \citep[e.g.,][]{Kovalev_2020}. These snapshot observations provided the angular size and \tb of the source by comparing the flux densities at ground-array and space baselines to a simple Gaussian model, if fringes were detected on the space baselines. The \oj survey observations were carried out during 2012$-$2018, and five experiments at L-band were conducted (2016 Apr.~1, 14, 18, 27, and 28) close to our observation (2016 Apr.~16). The survey observation results provided \tb$\sim2-4\times10^{12}$~K, which were well consistent with $T_{\rm B,min}$ at the ground-space baselines of our observation. At their baseline lengths which are longer than our measurement, the \tb were slightly lower than the extrapolated $T_{\rm B,min}$ from our observation. Note that the flux densities are mostly consistent in this observing period (see \autoref{fig:light_curve}) so the \tb difference may not be due to the flux variability but to the different ($u,v$) sampling of the five adjacent \ra survey observations towards PA$\sim-45^\circ$, where there is missing coverage in our imaging observation.  

The estimated \tb of C and J3 at 1.68\,GHz are $\approx10^{13}$~K and $\approx3\times10^{12}$~K, respectively. These are larger than both the inverse Compton limits of $(0.3-1.0)\times10^{12}$~K \citep{Kellermann_1969} and the equipartition limit of $\approx5\times10^{10}~$K \citep{Readhead_1994}, which can be explained at the inverse-Compton limit with a jet Doppler boosting factor, $\delta_{\rm j}\gtrsim3-10$. \citet{Gomez_2022} suggested the $\delta_{\rm j}\approx5-15$, based on the apparent jet speed, $\beta_{\rm app}$ $\leq15\,c$ at 15\,GHz \citep{Lister_2009, Lister_2016} and $\sim7-12\,c$ at 43\,GHz \citep{Weaver_2022}, so that the required $\delta_{\rm j}$ is in agreement with the independent kinematic constraints. With the same kinematic constraint, the \tb of J4, $\approx5\times10^{10}$~K, provides the intrinsic \tb of $\sim(0.3-1)\times10^{10}$~K that is slightly lower than the equipartition limit indicating a higher magnetic field energy density compared to the energy density of the relativistic particles \citep[e.g.,][]{Zhao_2022}. Note however that there are still considerable uncertainties of $\delta_{\rm j}$ and \tb measurements. 

\autoref{fig:btemp} shows the independent estimates of \tb, including the component model fitting (blue, horizontal lines), $T_{\rm B,min}$ and $T_{\rm B,lim}$ from visibility amplitudes (green and cyan squares), and the adjacent \ra snapshot survey measurements (yellow squares). Note that the \tb of component C corresponds to that of space baselines ($\gtrsim300$M$\lambda$) and is higher than maximum $T_{\rm B,lim}$ of ground baselines. This indicates inherent difficulty in finding this component with the ground-array at 1.68\,GHz. 
\\

\subsection{Polarization} \label{sec:polim} 

For the polarimetric analysis, the instrumental polarization leakage ($D-$term) was first calibrated using the Stokes~I image with the {\tt LPCAL} procedure in AIPS (see \autoref{fig:dterm} in \autoref{app:error}). Note that the group delay difference between RCP and LCP was already calibrated with the {\tt RLDLY}, as described in \autoref{sec:obs_calibration}. 
As a result, linear polarization is detected with a total fractional polarization of $m\sim$2.5\% and the EVPA is determined along with the structure (see \autoref{fig:image_I}). The EVPA was calibrated using a nearby single-dish observation with the EF at 2.64\,GHz on April~7, 2016, providing $m\sim3.89\%$ and EVPA$\sim-1\fdg06$ (F-GAMMA/QUIVER program; \citealt{Myserlis_2018, Angelakis_2019}). To account for the Galactic rotation measure (RM) of $31\pm3\,$rad~/~m$^{2}$ \citep{Rusk_1988} and the local RM in \oj of $-7.1\pm3.4\,$rad~/~m$^{2}$ \citep{Motter_2017}, EVPA rotations of $59\pm6^\circ$ and $-14\pm7^\circ$ were applied, respectively. 

The linear polarization properties of the model-fitted components are summarized in \autoref{tab:modelfit}. Note that the VLBI model components in total intensity and linear polarization do not correspond exactly to each other. Therefore, the linear polarization of each component is derived from the total intensity model in the image domain by masking the region outside of each model. 
The core region, C, displays low polarization ($m\sim0.1\%$), which is expected for optically thick emission. The EVPA of $\sim-$24$^\circ$ aligns well with the innermost jet structure of \oj at higher frequencies, indicating a dominant toroidal magnetic field \citep{Gomez_2022}. 
The J3 component also shows low polarization ($m\sim0.3\%$), with the EVPA slightly rotated by $\sim$10$^\circ$ towards the west compared to the core region. Lastly, the J4 component shows higher polarization ($m\sim$2.9\%) as it represents an optically thin jet component. The average EVPAs in these jet components are observed to be almost perpendicular to the local jet direction, which support the hypothesis that a poloidal magnetic field is dominant \citep{Myserlis_2018}. The uncertainties of the degree of polarization and EVPA are estimated based on the S/N of the total intensity and polarization images, as well as the standard deviation of the instrumental polarization amplitudes (see \citealt{Lico_2014} and \citealt{Gomez_2022} for more details). Additionally, for the EVPA, uncertainties arising from RM measurements have also been taken into consideration. 
\\


\section{Discussions and Conclusions} \label{sec:summary}

In this study, we have presented our findings on the total intensity and linear polarization structure of \oj at an observing frequency of 1.68\,GHz, obtained through the \ra observation on April~16, 2016. Our investigation fills a significant gap in the imaging attempts conducted at low frequencies (e.g., $\lesssim10\,$GHz), where limited exploration has occurred (\citealt{Perlman_1994, Fey_1996, Fey_1997}; see also the University College Cork Program\footnote{https://www.ucc.ie/en/raagn/restop/mojave/\#the-18-22cm-mojave-observations}). Notably, the use of the \srt, which possesses a baseline length of approximately $\sim5.6\,$\dearth, has enabled us to achieve the finest angular resolution image to date at this frequency. 

Our observation has unveiled the comprehensive jet structure oriented towards the south-west, accompanied by the identification of the optically thick core in its innermost region. This aligns well with the progressive bent-jet structure previously reported for \oj \citep[e.g.,][]{Hodgson_2017, Gomez_2022}. It is worth noting that the innermost jet structure towards the north-west, which has been observed at higher frequencies \citep[e.g.,][]{Gomez_2022, Lico_2022, Zhao_2022}, corresponds to the core region and remains unresolved in our study. By employing model fitting and examining the results up to $\sim3\,$mas (i.e., J4; see \autoref{tab:modelfit}), we have determined that the position angle (PA) of the jet, relative to the radio core, is $\sim-110^\circ$ (east of north). This is consistent with PA measurements made over the past $\sim30\,$years \citep[$\sim-112^\circ$;][]{Cohen_2017}. Especially, as the rotation of the inner jet \citep[e.g.,][]{Tateyama_2004} has changed from clockwise to counterclockwise in $\sim2010$, the jet ridgeline can clearly show the bending structure in following years. Therefore, our results support that the bending jet structure primarily arises from the rotation of the jet launching angle, which can also be amplified farther downstream due to the small viewing angle \citep{Gomez_2022}. Note that the origin of PA rotation could be the orbital motion of binary black hole system at the center \citep{Dey_2021}, although other scenarios cannot be ruled out, as introduced in \autoref{sec:intro}. 

By comparing the adjacent observations at 15, 43, and 86\,GHz, we have constructed a spectral index map that confirms the presence of both an opaque core and optically thin jet components. The model fitting has revealed three Gaussian components, corresponding to the core (C), brightest jet component at 1.68\,GHz (J3), and the extended faint component (J4). The spectral index of component C indicates that this is the optically thick core region, at least up to 28\,GHz, based on the SSA spectrum fitting. 
The sub-component C1 exhibits clearer transition from optically thick to thin across the frequency range spanning from 1.68\,GHz through 86\,GHz. This yields an estimated SSA turnover frequency of $\nu_\mathrm{m}\sim33\,$GHz, accompanied by a turnover flux density $S_\mathrm{m}\sim4.0\,$Jy (see \autoref{tab:spectral} and \autoref{fig:spix_comp}). Similarly, the sub-component C2 presents the opacity transition from observations up to 43\,GHz which provides the SSA turnover frequency and flux density as $\nu_\mathrm{m}\sim11.5\,$GHz and $S_\mathrm{m}\sim0.7\,$Jy, respectively.  

From the fitted $S_\mathrm{m}$ and $\nu_\mathrm{m}$, the magnetic field strength from SSA and assuming equipartition conditions can be estimated \citep[e.g.,][]{Pacholczyk_1970, Marscher_1983, Zdziarski_2014, Pushkarev_2019, Lee_2020, Gomez_2022}: 
\begin{equation}
 B_\mathrm{ssa} = 10^{-5} b(\alpha) \, \theta_\mathrm{m}^{4} \nu_\mathrm{m}^{5} S_\mathrm{m}^{-2}
 \dfrac{\delta_\mathrm{j}}{1+z} \, [\mathrm{G}]\,, 
\label{eq:b_SSA}
\end{equation}
\begin{equation}
B_\mathrm{eq} = 10^{-4} \left[\frac{(1+k_\mathrm{u})\, c_{12}\, \kappa_{\nu}}{f} \frac{S_\mathrm{m}\, \nu_\mathrm{m}}{\theta^{3}_\mathrm{m}\,D_\mathrm{L,Gpc}} 
\frac{(1+z)^{10}}{\delta_\mathrm{j}^4} \right]^{2/7} \, [\mathrm{G}]\,. 
\label{eq:b_eq}
\end{equation}
In \autoref{eq:b_SSA}, $b\left(\alpha\right)$ is a coefficient as a function of the spectral index \citep[e.g.,][]{Marscher_1983, Pushkarev_2019}, $\theta_\mathrm{m}$ is the emitting region diameter in mas, and $\delta_\mathrm{j}$ is the Doppler boosting factor. Note that a factor of 1.8 is multiplied to $\theta_\mathrm{m}$ to convert the Gaussian to the spherical shape before estimating the magnetic field strengths \citep{Marscher_1983}. 
In \autoref{eq:b_eq}, $f$ is volume filling factor of the emitting plasma, $D_\mathrm{L,Gpc}$ is the luminosity distance in Gpc, $k_\mathrm{u}$ is the ratio of total energy to the SSA population of electrons in the emitting region, $c_{12}$ is a coefficient \citep[e.g., ][]{Pacholczyk_1970}, and $\kappa_\nu$ is a function of spectral index and frequency cutoffs of the SSA emission. 
In this study, we adapted the same assumption with \citet{Gomez_2022}: $f=1$, $k_\mathrm{u}=1$ (i.e., an electron-positron flow), $\kappa_{\nu} \approx (\nu_\mathrm{max}/\nu_\mathrm{m})^{1+\alpha}/(1+\alpha)$ where $\nu_\mathrm{max}=10\,$THz. 

With $\delta_{\rm j}=10$ \citep[e.g.,][]{Gomez_2022}, the magnetic field strength of component C1 is estimated as $B_{\rm ssa}\sim3.4\pm1.4\,$G and $B_{\rm eq}\sim2.6\pm0.2\,$G. As for the component C2, $B_{\rm ssa}\sim1.0\pm0.8\,$G and $B_{\rm eq}\sim1.6\pm0.3\,$G are obtained. Components J3 and J4 are considered optically thin jet components, and their SSA parameters have not been well constrained. Note that even if C and J3 exhibit $T_{\rm B}$ exceeding the inverse Compton limit, these can still be explained by a jet Doppler boosting factor of $\delta_{\rm j}\approx5-15$, which would be in agreement with kinematic constraints at higher frequencies. 

The integrated degree of linear polarization is found as $\sim$2.5\%. As expected, low polarization of $\sim0.1$\% is obtained in the opaque core region, while higher polarization levels of up to $\sim$3\% are observed in more distant jet components. The EVPA of linear polarization is well consistent with the local jet direction in the core region, which may support a toroidal magnetic field structure \citep{Gomez_2022}. 
Farther downstream of the jet, on the other hand, the EVPAs are almost perpendicular to the jet, indicating a predominant poloidal magnetic field structure \citep{Myserlis_2018}. However, it is important to note that significant uncertainty remains in the alignment between the local jet axis and the EVPA. This is due to the limited angular resolution, which challenges the determination of a reliable jet ridge line, as well as the substantial error and potential time variation in the RM. 

\oj represents a prime candidate for hosting a supermassive black hole binary system, from which a jet from the secondary black hole has been also predicted \citep[e.g.,][]{Dey_2021}. Despite efforts by \citet{Zhao_2022} to search for a secondary jet from the predicted secondary SMBH using GMVA+ALMA observations at 86 GHz, no conclusive detection has been made, potentially due to sensitivity limitations. Similarly, our study at 1.68\,GHz, even with the space-ground VLBI observation using \ra, has failed to detect the secondary jet. Therefore, this will be further studied, for instance with the EHT at 230\,GHz (J.~L.~G{\'o}mez et al.~in prep.) and with  \ra at 22\,GHz (T.~Traianou et al.~in prep.; R.~Lico et al.~in prep.). 
\\

\begin{acknowledgements}
We are grateful to anonymous referee, B.~Boccardi, T.~Savolainen, and J.~Park for their helpful comments on the manuscript.  
The work at the IAA-CSIC is supported in part by the Spanish Ministerio de Econom{\'i}a y Competitividad (grants AYA2016-80889-P and PID2019-108995GB-C21), the Consejería de Econom{\'i}a, Conocimiento, Empresas y Universidad of the Junta de Andaluc{\'i}a (grant P18-FR-1769), the Consejo Superior de Investigaciones Cient{\'i}ficas (grant 2019AEP112), and the grant CEX2021-001131-S funded by MCIN/AEI/ 10.13039/501100011033. 
This publication acknowledges the project M2FINDERS, which is funded by the European Research Council (ERC) under the European Union's Horizon 2020 research and innovation program (grant agreement No.~101018682). 
\rev{IC is supported by the KASI-Yonsei Postdoctoral Fellowship.} 
FMP acknowledges support by the ERC under the HORIZON ERC Grants 2021 programme under grant agreement No. 101040021. 
JYK is supported for this research by the National Research Foundation of Korea (NRF) grant funded by the Korean government (Ministry of Science and ICT; grant no.~2022R1C1C1005255). 
%
The \ra project is led by the Astro Space Center of the Lebedev Physical Institute of the Russian Academy of Sciences and the Lavochkin Scientific and Production Association under a contract with the Russian Federal Space Agency, in collaboration with partner organizations in Russia and other countries. 
The European VLBI Network is a joint facility of independent European, African, Asian, and North American radio astronomy institutes. 
The VLBA is an instrument of the National Radio Astronomy Observatory, a facility of the National Science Foundation operated under cooperative agreement by Associated Universities. 
This research is based on observations correlated at the Bonn Correlator, jointly operated by the Max-Planck Institut f{\"u}r Radioastronomie (MPIfR), and the Federal Agency for Cartography and Geodesy (BKG). 
This research has made use of data from the MOJAVE database that is maintained by the MOJAVE team \citep{Lister_2018}. 
This study makes use of 43\,GHz VLBA data from the VLBA-BU Blazar Monitoring Program (VLBA-BU-BLAZAR; \url{http://www.bu.edu/blazars/VLBAproject.html}), funded by NASA through the Fermi Guest Investigator grant 80NSSC20K1567. 
For data correlation, calibration, imaging and analysis, the softwares of {\tt DiFX} \citep{Bruni_2016}, {\tt AIPS} \citep{Greisen_2003}, {\tt DIFMAP} \citep{Shepherd_1994}, and {\tt eht-imaging} \citep{Chael_2018} are used, respectively. 
\end{acknowledgements}


\bibliography{oj287}{}
\bibliographystyle{aa}


\newpage
\begin{appendix}

\section{Data issue, image fidelity, and instrumental polarization} 
\label{app:error}

\begin{figure*}[t]
\centering 
\includegraphics[width=\linewidth]{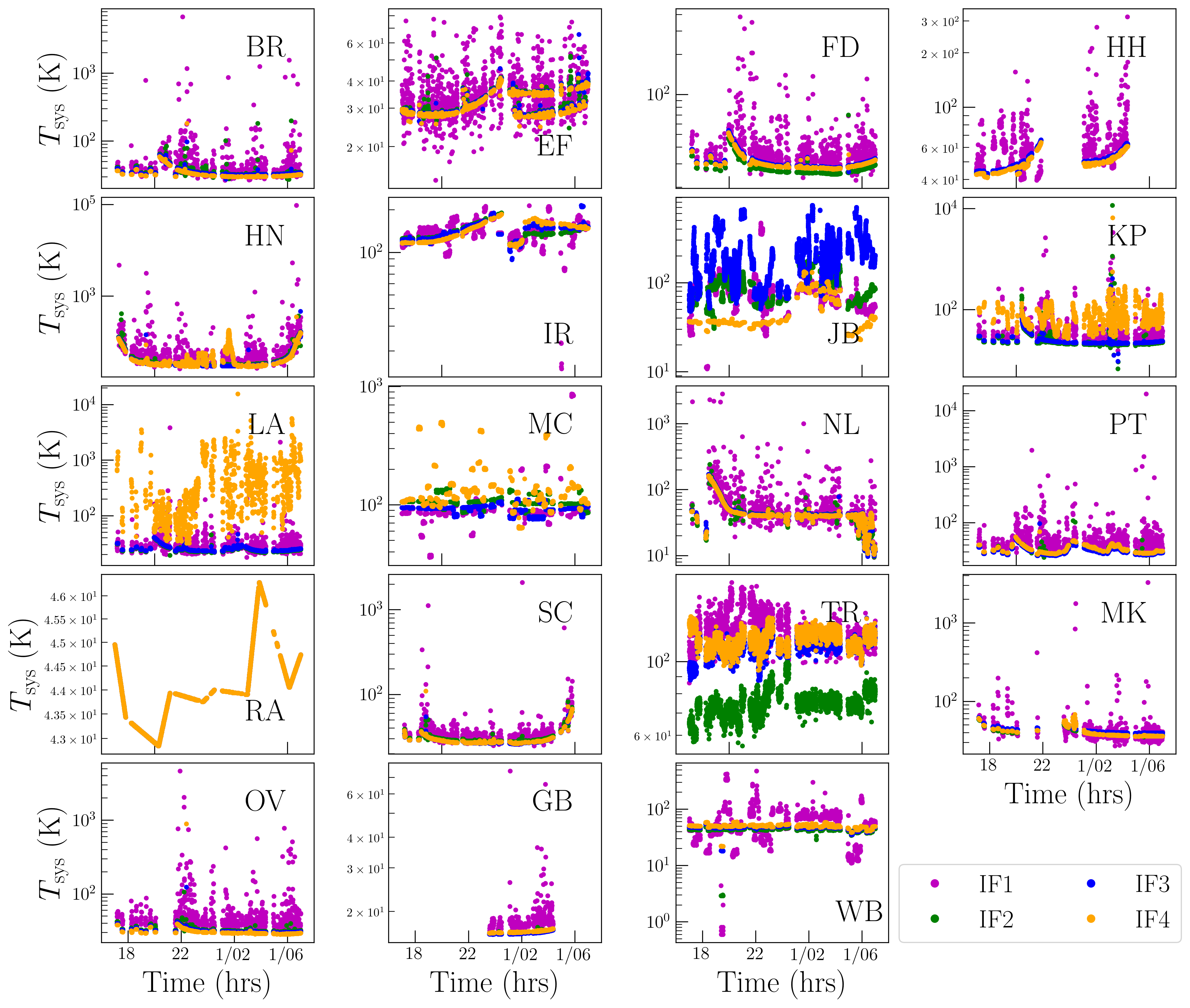}  
\caption{
The system temperature, $T_{\rm sys}$, at all stations for LCP. Each IF is shown with different colors: IF1 (magenta), IF2 (green), IF3 (blue), and IF4 (orange). Note that there is no $T_{\rm sys}$ measurements at RO so it has been excluded. 
}
\label{fig:tsys}
\end{figure*}

\begin{figure*}[t]
\centering 
\includegraphics[width=0.45\linewidth]{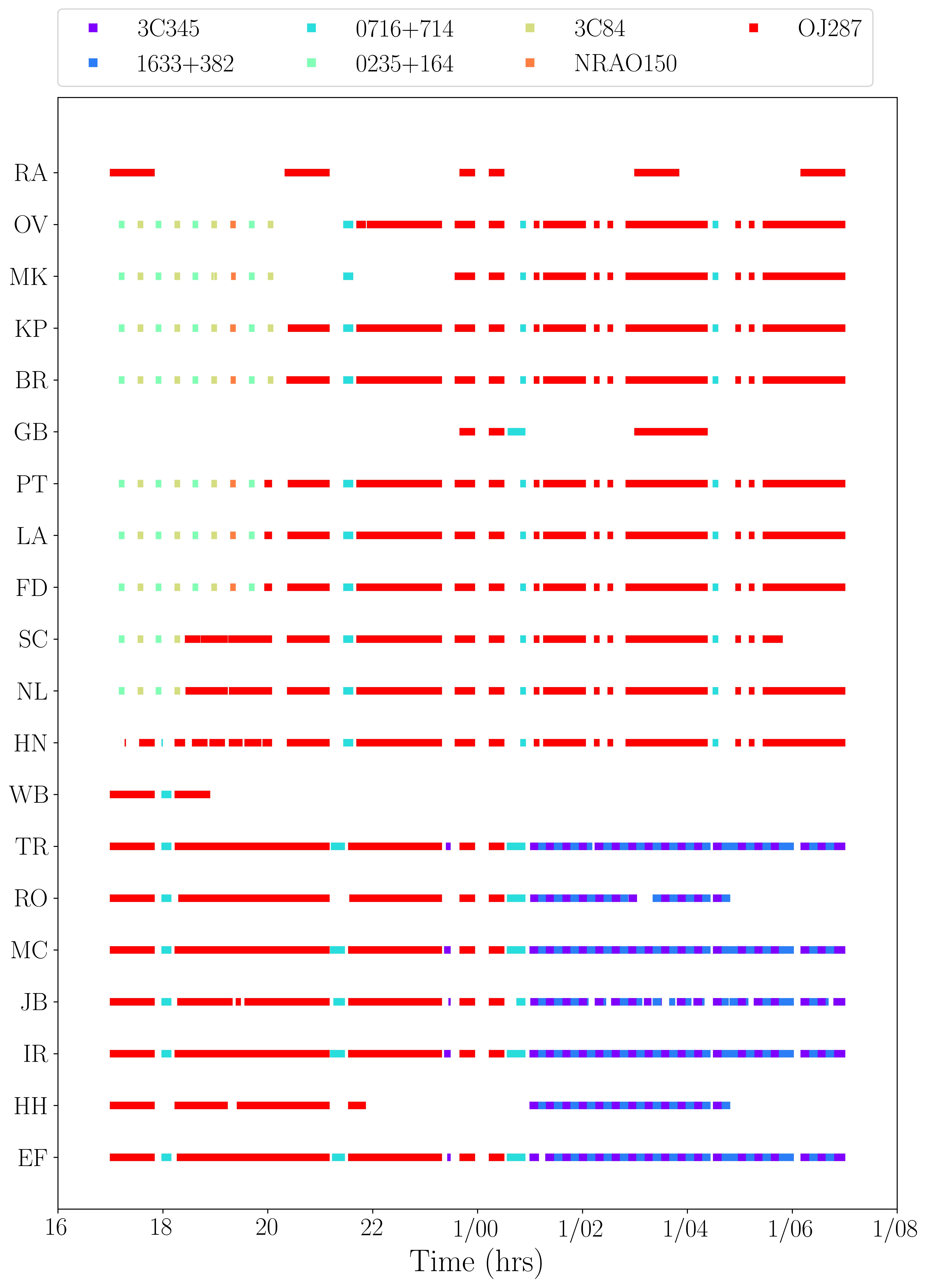}  
\includegraphics[width=0.54\linewidth]{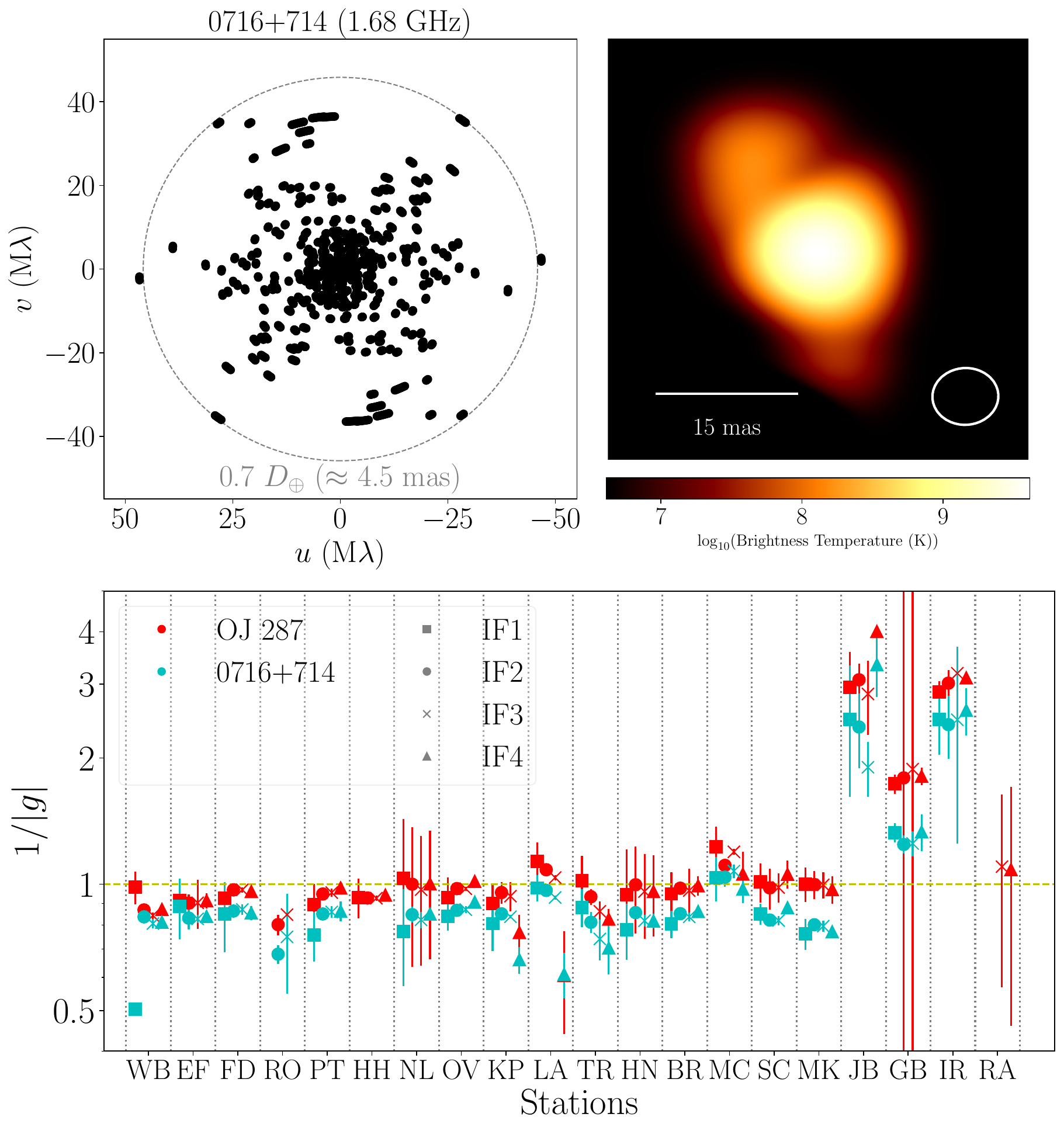} 
\caption{
(Left) Sources observed at each station, as a function of observing time. 
\oj is shown with red color, and the other calibrators are also shown with different colors: 3C\,345, 1633$+$382, 0716$+$714, 0235$+$164, 3C\,84, and NRAO\,150. 
Among them, the 0716$+$714 is suitable to compare the station-based gain correction with the \oj, thanks to its station coverage. 
(Right, top) The ($u,v$)-coverage towards 0716$+$714 and its image are shown. 
Note that the \srt has not observed this source and the longest baseline length is limited up to $\sim50\,$M$\lambda$ so the angular resolution ($\sim4.5\,$mas) is much poorer than for \oj. 
(Right, bottom) The multiplicative gain correction factors at each station, derived from \oj (red) and 0716$+$714 (cyan). 
Different markers indicate each IF from 1 to 4 at each station. 
}
\label{fig:gain}
\end{figure*}

In our observation, certain stations and IFs were flagged out due to the poor data quality. First, spurious amplitudes are present at 1) IF1 (1.620-1.636\,GHz), 2) all IFs at JB, and 3) IF4 (1.668-1.684\,GHz) at KP, LA, and MC. These are also present in the $T_{\rm sys}$ measurements (see \autoref{fig:tsys}) implying that it is a station-based issue. Note that the large scatter at IF1 may be due to radio interference, since it is affecting all the stations. These data were initially excluded from imaging to avoid any possible bias during the image reconstruction. Once an initial image was obtained these data were self-calibrated in amplitude with that image in an attempt to correct for the issues mentioned before. The same procedure was used for RO, GB, and IR for which only single polarization data could be used. These provide the larger gain correction factor, as shown in \autoref{fig:gain} (lower, right), in comparison with the results from a calibrator 0716$+$714. \autoref{fig:gain} shows the observed sources at each station, as a function of total observing time. To compare the gain correction at each station, 0716$+$714 was the most suitable calibrator as it was observed across all participating stations except \srt. 
In this regard, the ($u,v$)-coverage towards 0716$+$714 and its image are shown together. As a result, it is found that the multiplicative gain correction factor, $1/|g|$, is well consistent in \oj and 0716$+$714. In addition, as expected, larger $1/|g|$ are derived from the flagged data. 

\begin{figure*}[t]
\centering 
\includegraphics[width=0.49\linewidth]{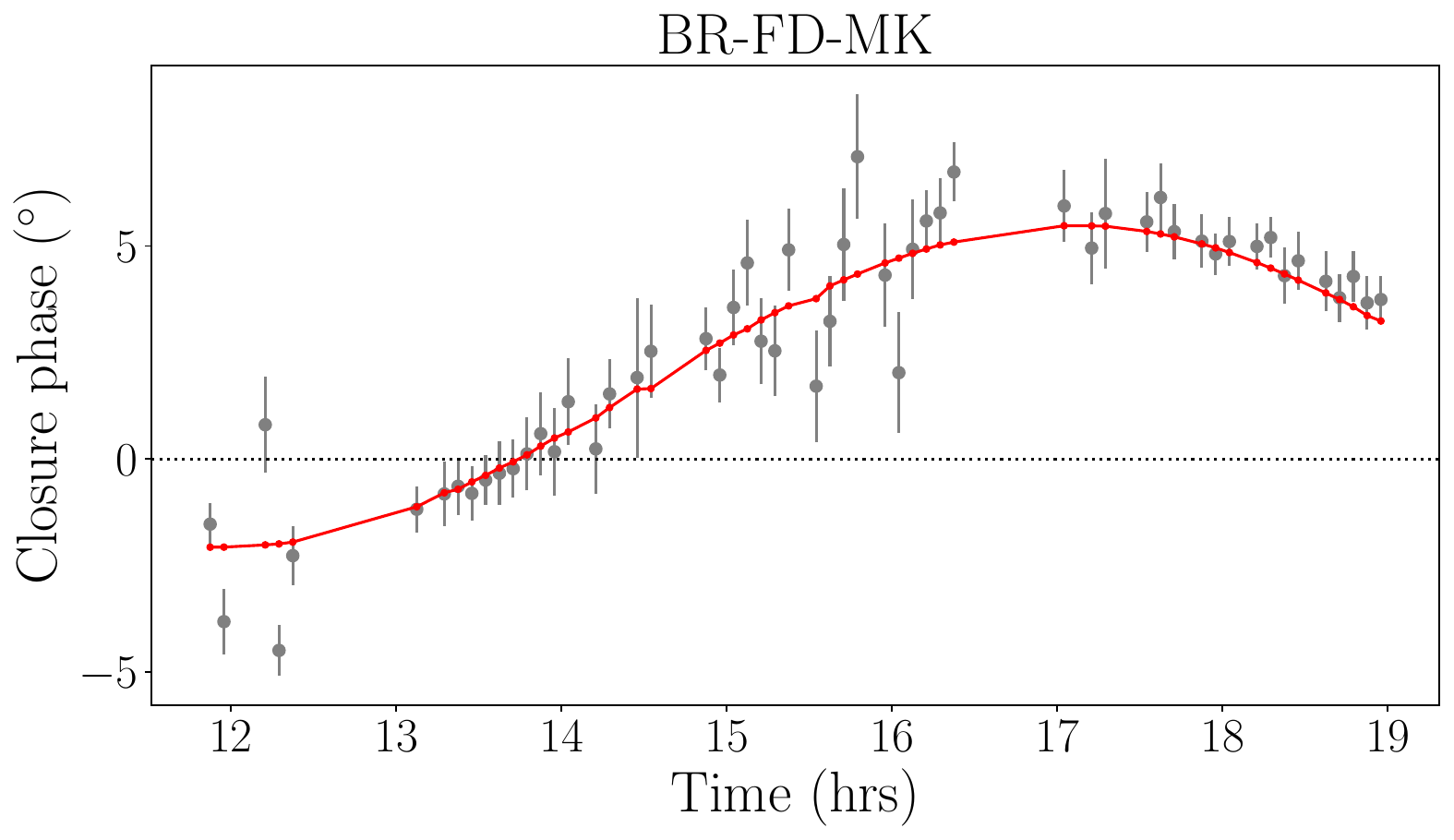}
\includegraphics[width=0.49\linewidth]{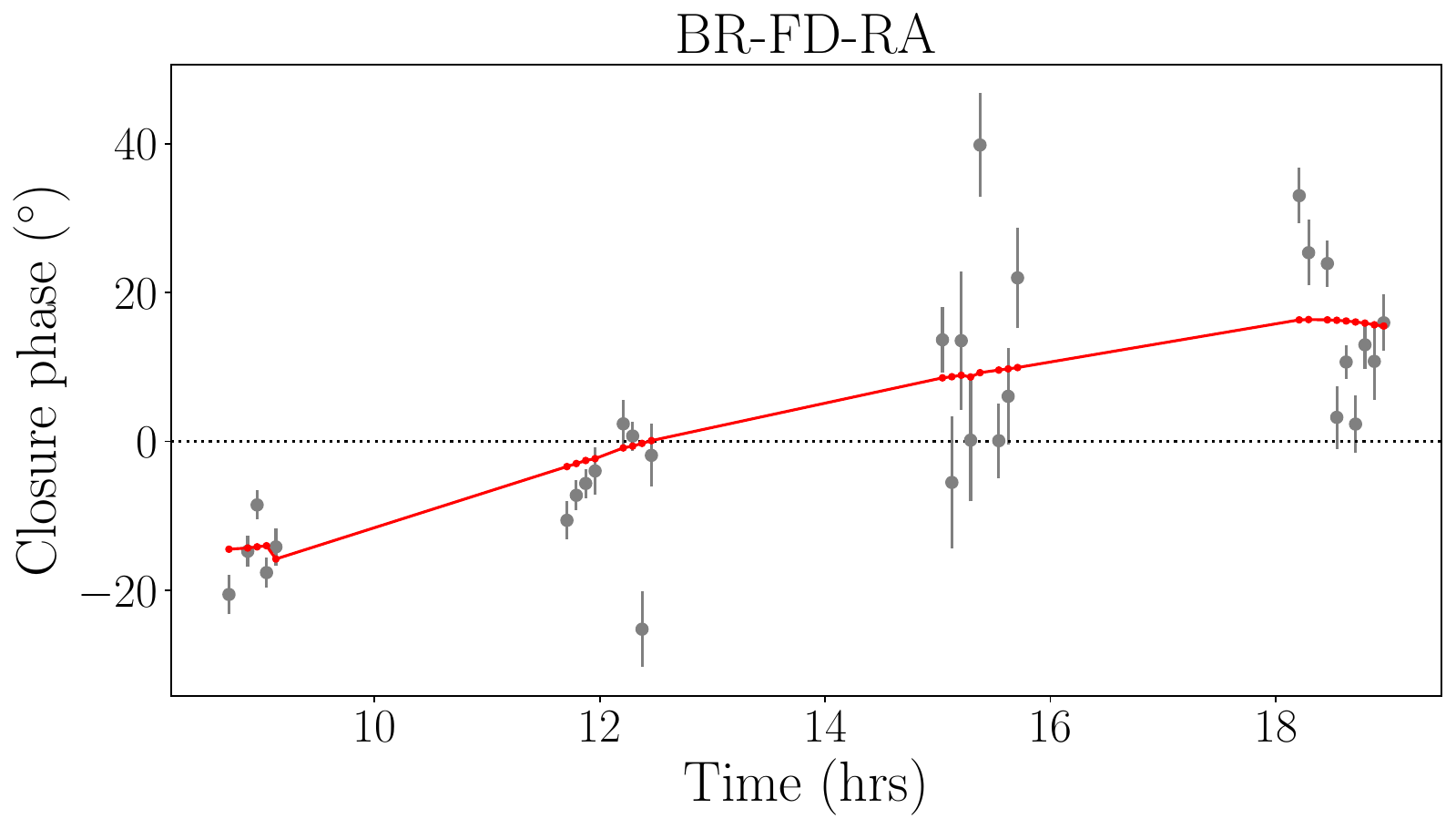}
\includegraphics[width=0.49\linewidth]{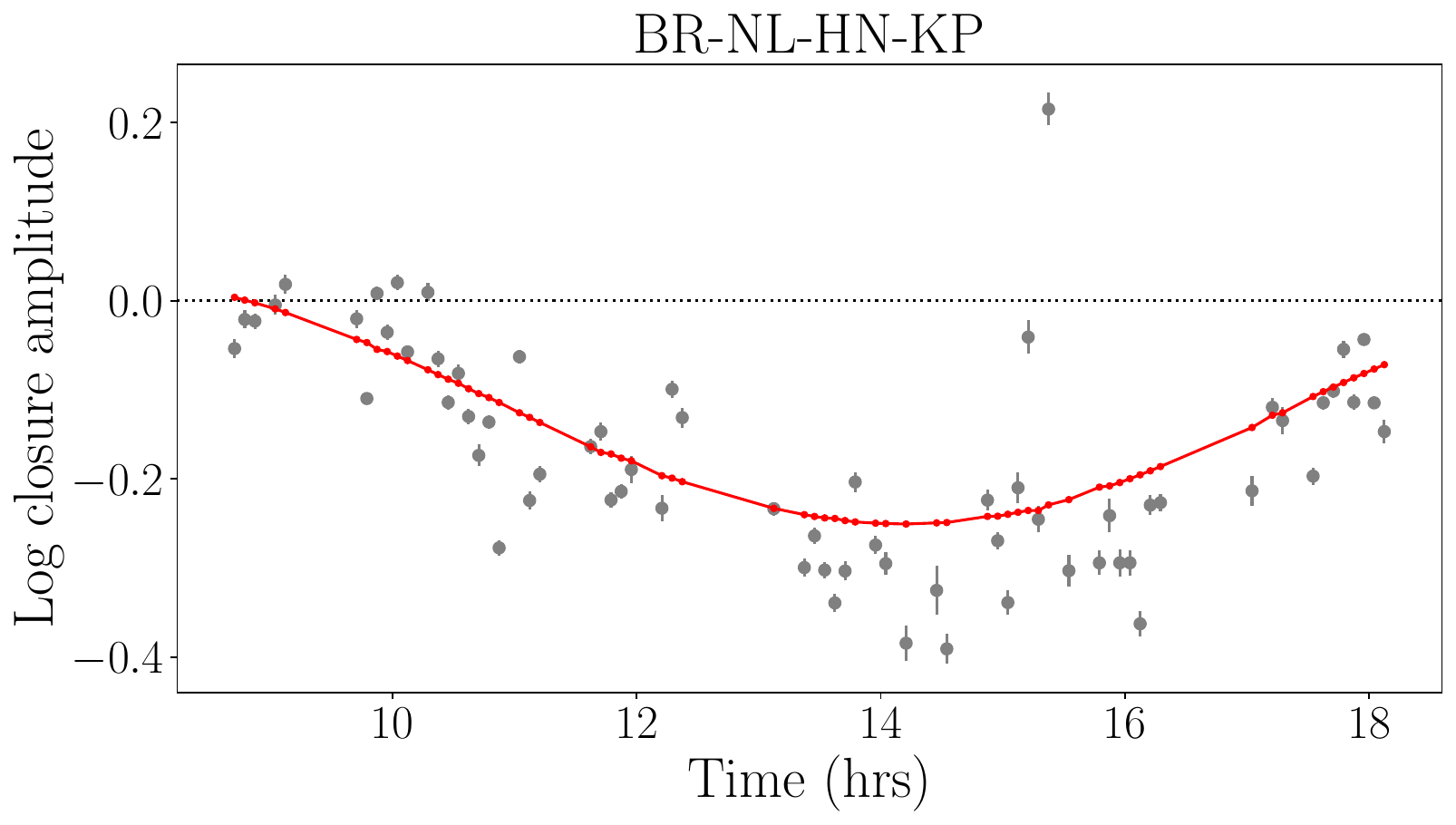}
\includegraphics[width=0.49\linewidth]{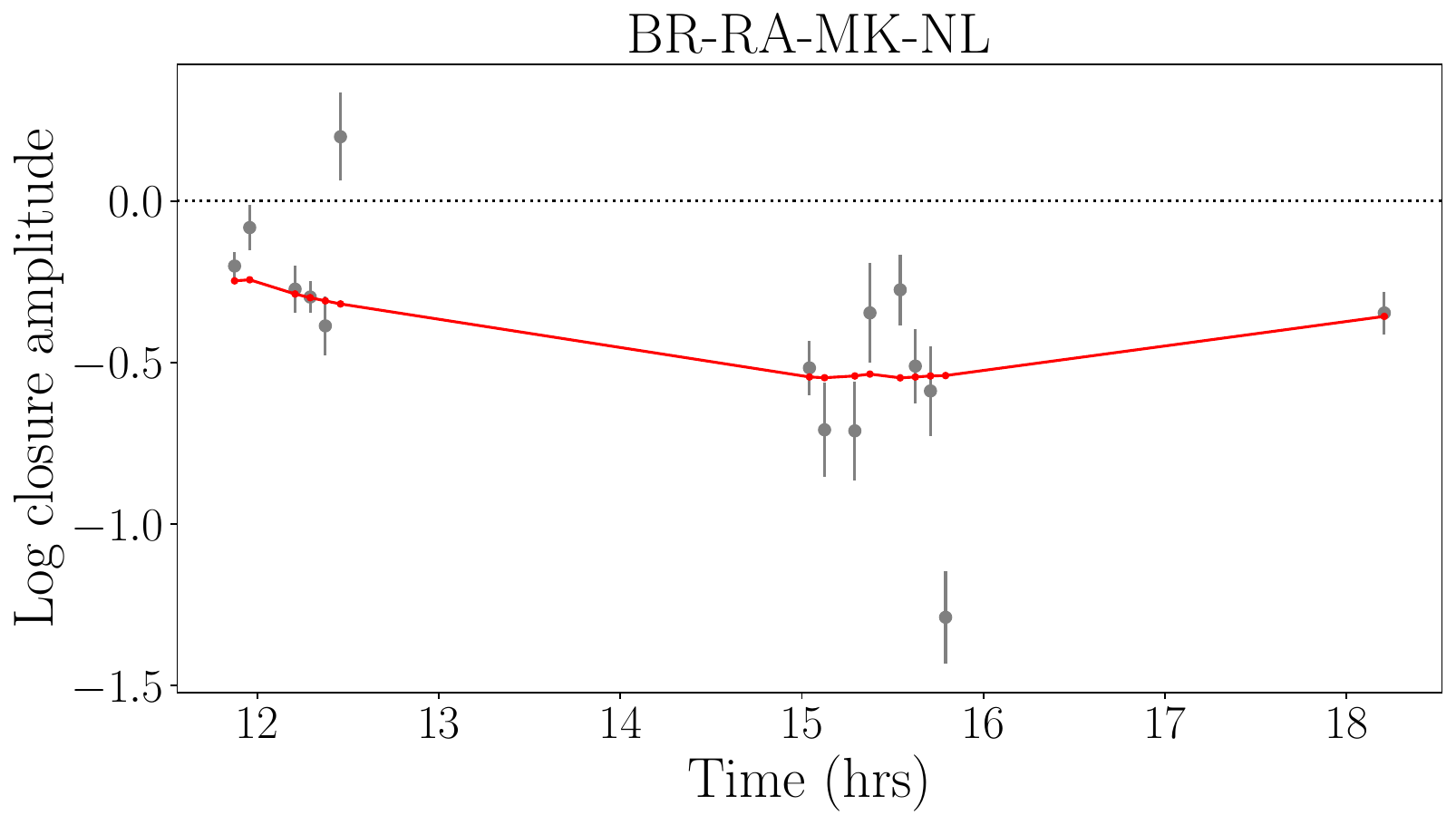}
\caption{
The closure phases (top) and log closure amplitudes (bottom), at the triangle/quadangle of ground array (left) and with the \srt (right). The stations consisting of each triangle and quadangle are shown at the top of each panel. Model from image (red, solid line) well fits the data (gray, circle). 
}
\label{fig:closure_fit}
\end{figure*}

Based on the data, the \difmap imaging was implemented with different parameters. The best image has been selected based on the $\chi_{\nu}^{2}$ of closure quantities. \autoref{fig:closure_fit} shows the example closure phases (upper) and log closure amplitudes (lower) of observed data (gray points) and the model (red, solid line). Note that this shows two representative triangles / quadangles with ground array (left) and with \srt (right). 
As shown, the model well fits the data with $\chi_{\nu}^{2}\approx1.7$ and $\approx2.2$ for closure phase and log closure amplitude respectively. 

\begin{figure*}[t]
\centering 
\includegraphics[width=\linewidth]{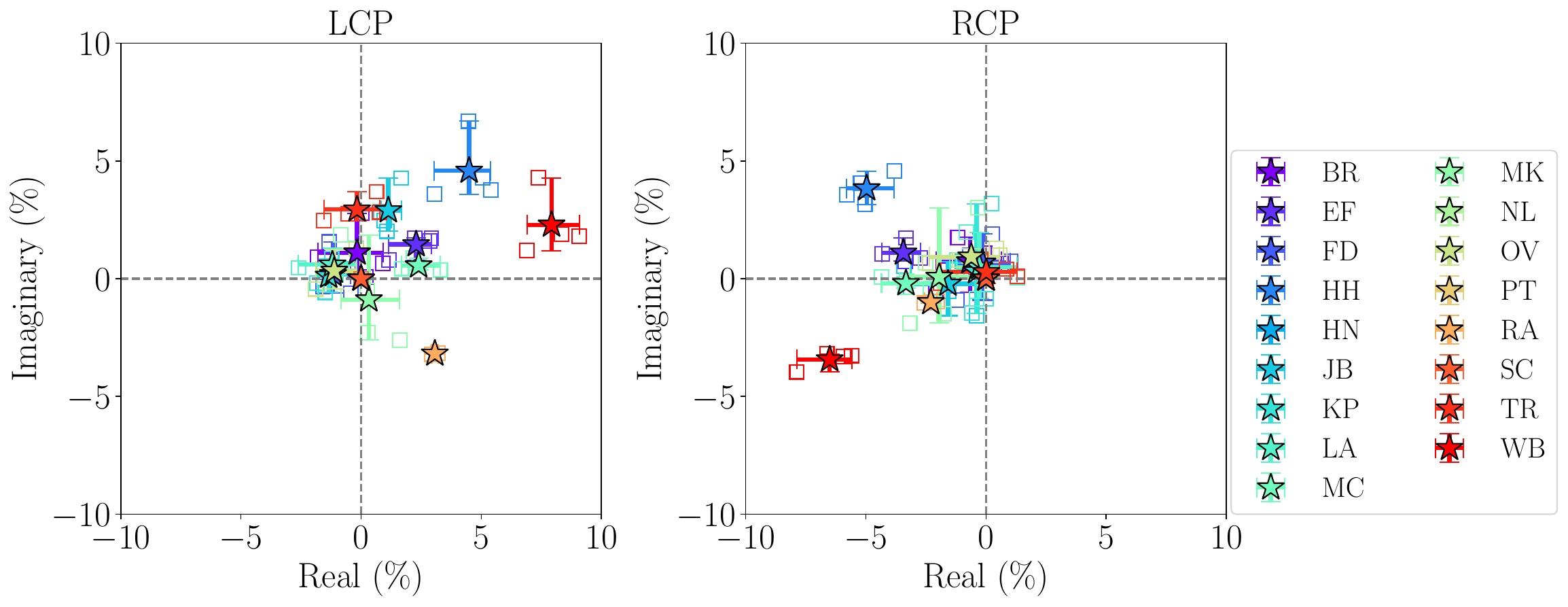}
\caption{
Derived instrumental polarization, D-term, in \%. The left and right panels show the solutions in LCP and RCP, respectively. D-terms at each antenna are shown with different colors, and the squares show the solution at each IF. The asterisk with error bar is the averaged D-term across IFs. 
}
\label{fig:dterm}
\end{figure*}

Lastly, the calibration of the instrumental polarization leakage, so called D-term, is shown in \autoref{fig:dterm}. This is required to obtain reliable polarimetric results. To derive the D-term, the Stokes~I data are first self-calibrated with the CLEAN image, as described in \autoref{sec:obs_calibration}. Then the self-calibrated data and image are loaded to AIPS, and the cross polarizations (i.e., LR and RL) are further self-calibrated using {\tt CALIB}. 
After that, the polarization structure of \oj is first specified with {\tt CCEDT} and the D-term calibration is applied with {\tt LPCAL}. This is because {\tt LPCAL} assumes that the source consists of sub-components having constant fractional polarization in each. Note that the D-term solution is found for each IF separately, so \autoref{fig:dterm} shows all these values (square) together with the averaged one (asterisk). The error bar of averaged value indicates the dispersion range of the D-term across different IFs, and the results for LCP and RCP are shown in left- and right-panel respectively.

\end{appendix}

%


\end{document}